\documentclass[12pt]{article}
\usepackage{graphicx}
\usepackage{amsfonts}
\usepackage{amsmath}
\usepackage{amstext}
\usepackage{tabularx}
\usepackage{mathrsfs}
\usepackage{natbib}
\usepackage{setspace}
\usepackage{amssymb}
\usepackage{color}
\usepackage{rotating}
\usepackage{multirow}
\usepackage{amsmath,amssymb,amsthm}
\usepackage{natbib}
\usepackage{booktabs}
\usepackage{bm}
\usepackage{thrmappendix}
\usepackage{threeparttable}
\usepackage{longtable}

\usepackage{fancyvrb}

\usepackage[utf8]{inputenc}

\usepackage{pdflscape}

\usepackage[shortlabels]{enumitem}

\theoremstyle{plain}
\newtheorem{theorem}{Theorem}

\newtheorem{lemma}{Lemma}

\theoremstyle{definition}

\newtheorem{remark}{Remark}

\DeclareMathOperator{\Var}{Var}
\DeclareMathOperator*{\argmin}{arg\,min}

\newcommand{\pconv}{\xrightarrow{p}}

\newcommand{\tr}{^{\top}}
\newcommand{\pconvs}{\xrightarrow{p^\ast}}
\newcommand{\dconvs}{\xrightarrow{d^\ast}}
\newcommand{\ud}{\textnormal{d}}

\def\E{{\textnormal{E}}}

\def\R{{\mathbb{R}}}
\def\P{{\textnormal{P}}}
\def\1{{\mathbf{1}}}
\def\calL{\mathcal{L}}

\begin{document}

\title{Bootstrap inference for panel data quantile regression\footnote{The authors would like to express their appreciation to Hide Ichimura, Roger Koenker, and participants in the seminars at University of Arizona and the 26th International Panel Data Conference for helpful comments and discussions. Computer programs to replicate the numerical analyses are available from the authors.  All the remaining errors are ours.}}
\author{Antonio F. Galvao\thanks{%
%Corresponding Author. 
Department of Economics, Michigan State University, East Lansing, USA. E-mail: \texttt{agalvao@msu.edu}}\\
\and Thomas Parker\thanks{%
Department of Economics, University of Waterloo, Waterloo, Canada. E-mail: \texttt{tmparker@uwaterloo.ca}}\\
\and Zhijie Xiao\thanks{%
Department of Economics, Boston College, Chestnut Hill, USA. E-mail: \texttt{xiaoz@bc.edu}}\\
}

\maketitle

\begin{abstract}
\begin{spacing}{1}
This paper develops bootstrap methods for practical statistical inference in panel data quantile regression models with fixed effects. We consider random-weighted bootstrap resampling and formally establish its validity for asymptotic inference. The bootstrap algorithm is simple to implement in practice by using a weighted quantile regression estimation for fixed effects panel data. We provide results under conditions that allow for temporal dependence of observations within individuals, thus encompassing a large class of possible empirical applications. %Monte Carlo simulations provide numerical evidence that confidence intervals based on the asymptotic normal approximation can be very distorted in finite samples. Instead, the proposed bootstrap greatly reduces these distortions, and the simulations confirm that the proposed methods have correct finite sample properties. 
Monte Carlo simulations provide numerical evidence the proposed bootstrap methods have correct finite sample properties. 
Finally, we provide an empirical illustration using the environmental Kuznets curve.
\end{spacing}
\end{abstract}
\thispagestyle{empty}



%\newpage
%%
%%
%{%\baselineskip=0%-.001in 
%\parskip -.01in
% \tableofcontents  }
%  \thispagestyle{empty}
%  \newpage
%%  \listofchanges[style=<list|summary>]
%% \newpage

\newpage

%\doublespacing

\section{Introduction}

The quantile regression (QR) model has been widely used to to capture the heterogeneous effects that covariates may have on an outcome of interest, allowing the analyst to investigate a wide variety of forms of conditional heterogeneity under weak distributional assumptions. In program evaluation studies in economics, finance, and statistics, conditional quantile methods help to analyze how a treatment or social program affects the entire outcome distribution of interest.
Moreover, since the work of \cite{Matzkin03} and \cite{deCastroGalvao19}, QR has been used for empirical structural work and can provide a natural way to represent structural relationships.

\cite{Koenker04} introduced a general approach for estimation of QR panel models with individual specific fixed effects. Recently, there has been a growing literature on estimation and inference for panel data fixed effects (FE) QR models (FE-QR). The FE-QR model is designed to control for individual specific heterogeneity while exploring heterogeneous covariate effects, and therefore provides a flexible method for the analysis of panel data models.\footnote{For other recent developments, see, among many others, \cite{Canay11}, \cite{KatoGalvaoMontes-Rojas12},  \cite{GalvaoLamarcheLima13}, \cite{ChernozhukovFernandezValHahnNewey13}, \cite{GalvaoWang15},  \cite{ArellanoBonhomme16}, \cite{ChetverikovLarsenPalmer13}, \cite{GalvaoKato17}, \cite{GrahamHahnPoirierPowell2015}, \cite{GuVolgushev19}, \cite{MachadoSantosSilva19}, and \cite{ZhangWangZhu19}.}

Inference in panel data quantile methods with FE has relied mainly on asymptotic approximations for computation of the variances of estimators. These variances depend on the conditional density of the innovations, which can be difficult to compute in practice. %\textcolor{red}{(DELETE THIS SENTENCE?)} 
We provide simulation evidence that confidence intervals for the FE-QR estimator based on the asymptotic normal approximation can be distorted in finite samples.  The bootstrap has the advantage that it does not require estimating the variance-covariance matrix, which allows one to circumvent the issue of bandwidth selection for density estimation.  The use of the bootstrap as an alternative to asymptotic approximations has been informally considered, but its formal properties have not received the same amount of attention as in the cross-section or time series QR literatures. In the QR panel data FE context, \cite{AbrevayaDahl08}, \cite{Lamarche08}, \cite{Canay11},  \cite{GalvaoMontesRojas15}, and \cite{GrahamHahnPoirierPowell2015} use bootstrap for constructing confidence intervals, but do not provide formal theoretical validity for the procedures they use, nor do they provide an explicit methodology to implement the bootstrapping for FE-QR panel data.  \citet{LamarcheParker21} consider wild bootstrap for penalized QR and FE-QR estimators with panel data.

We contribute to this literature by formalizing the properties of bootstrap inference methods for QR panel data models with FE. We consider a random-weighted bootstrap (see, e.g., \cite{RaoZhao92}, \cite{PraestgaardWellner93}, and \cite{ShaoTu95}) and show that it can be used to construct asymptotically valid bootstrap standard errors, confidence intervals, and hypothesis tests for the parameters of interest. We formally establish the asymptotic validity of the random-weighted bootstrap by showing distributional consistency of the bootstrap method.  This result implies consistency of bootstrap percentile confidence intervals. Nevertheless, it does not necessarily imply consistency of the bootstrap variance-covariance estimation. Hence, we also prove consistency of the resampled variance-covariance matrix as an estimator for the asymptotic covariance matrix. Given the consistency of these bootstrap procedures, the practical construction of standard errors, confidence intervals, and hypothesis tests are very simple and based on the empirical distribution of the bootstrapped regression coefficients of interest. The weights are only a function of the cross-section dimension, meaning that serial correlation in the observations will be preserved in the bootstrap data generating process.\footnote{This method is applicable to a wide variety of models and not limited to independent and identically distributed ($i.i.d.$) errors in regression models.  Specifically, we show that this bootstrap is generally consistent for observations that are $\beta$-mixing in the time dimension, using nearly the same conditions that \citet{KatoGalvaoMontes-Rojas12} and \citet{GalvaoGuVolgushev20} used to show asymptotic normality of FE-QR slope coefficient estimates.}

A prominent challenge to the use of FE-QR models, and associated asymptotic methods, is that we have to deal with the incidental parameters problem even for a linear model.\footnote{The incidental parameters problem here means that as the sample size grows the number of parameters, i.e., the number of individual fixed effects, increase as well. In panel QR models there is no simple transformation that removes FE and at the same time preserves the functional form of a structural equation.} To overcome this drawback, we use recent advances in the panel QR literature \citep{GalvaoGuVolgushev20} and deal with the incidental parameters problem explicitly.\footnote{\cite{GalvaoGuVolgushev20} derive the asymptotic normality of the FE-QR estimator for large panels under a relatively weak condition on the sample size growth, specifically $n(\log T)^{4}/T \to 0$ under temporally dependent observations. This significantly improves upon previous conditions in the literature. A key insight was a detailed analysis of the expected values of remainder terms in the classical Bahadur representation for QR, while previous approaches focused on the stochastic order of those remainder terms. A similar analysis was previously performed in \cite{VolgushevChaoCheng19} for general QR models with growing dimension under the assumption of independent and identically distributed observations.}  Moreover, we highlight that these results apply under temporal dependence of observations within individuals, and thus encompass a large class of possible empirical applications.

%In particular, we derive the asymptotic results under an assumption that the length of time, $T$, and the number of individuals (groups), $N$, grow simultaneously. A key insight in the proofs is a detailed analysis of the expected values of remainder terms in the classical Bahadur representation for QR, while previous approaches (including Kato, Galvao, and Montes-Rojas (2012) and Galvao and Wang (2015)) focused on the stochastic order of those remainder terms. A similar analysis was previously performed in Volgushev, Chao, and Cheng (2018) for general QR models with growing dimension under the assumption of independent and identically distributed observations. The proofs involve subtle empirical process arguments, and extending those results to settings with dependent data requires a substantial amount of work. 

The algorithm for practical implementation of the random-weighted bootstrap method is simple. First, draw non-negative $i.i.d.$ random weights from an appropriate distribution satisfying mean and variance both equal to one. Second, for a given quantile of interest, fit a weighted panel FE-QR model, weighing unit $i$'s contribution to the objective function with the $i$th weight generated from the first step. These first and second steps are repeated many times. Finally, compute percentile confidence interval from the empirical distribution of bootstrapped coefficients, or calculate the variance-covariance matrix of the bootstrap coefficients to construct confidence intervals or Wald-type test statistics.

A Monte Carlo study is conducted to evaluate the finite sample properties of the proposed bootstrap inference procedure. We use the FE-QR estimator. We evaluate the bootstrap procedure using an $i.i.d.$ and a dependent case. We compute empirical coverage rates for confidence intervals using the bootstrap procedures. The results show evidence empirical coverage for confidence intervals based on the proposed bootstrap are close to the nominal size, and as expected, the results indicate that coverage improves as the sample size increases.  Furthermore, this bootstrap performs well as a variance estimator under nearly the same regularity conditions as those required for bootstrap distribution consistency.  Overall, the results show good performance for relatively small samples.

In an empirical example we examine the relationship between CO$_{2}$ emissions and economic development. In particular, we examine the Environmental Kuznets Curve (EKC) hypothesis for a panel of 24 OECD countries and 32 non-OECD countries using a panel data QR model which allows us to account for heterogeneous effects. The EKC hypothesis describes an inverted U-shaped relationship between income inequality and income per capita. The results corroborate the nature and validity of the income-pollution relationship based on the EKC hypothesis for OECD countries. Therefore, the results show evidence that the monotonically increasing relationship across the conditional distribution of CO$_{2}$ emissions relates to the level of economic development of the country. For the non-OEDC countries, although we also find empirical evidence such a income-pollution relationship, there is no evidence of validity of the EKC hypothesis.

Now we briefly review the literature related to this paper. There are two different branches. First, bootstrap techniques have been used to construct confidence intervals for QR models in the cross-sectional context extensively. \cite{Buchinsky95} uses Monte Carlo simulation to study and compare several estimation procedures of the asymptotic covariance matrix in QR models, and the results favor the pairwise bootstrap design. \cite{Hahn95} shows that the construction of confidence intervals based for the QR estimators can be greatly simplified by using bootstrap. Moreover, the confidence intervals constructed by the bootstrap percentile method have asymptotically correct coverage probabilities. \citet{BoseChatterjee03} show consistency of multiplier bootstrap methods for quantile regression (among other estimators that are found by minimizing a convex objective function).  \cite{Horowitz98} proposes the smoothed least absolute deviation (LAD) estimator and shows that the bootstrap provides asymptotic refinements for hypothesis tests and confidence intervals based on the smoothed LAD estimator.  \cite{FengHeHu11} propose an adaptation wild residual bootstrap methods for QR. \cite{WangHe07} develop inference procedures based on rank-score tests with random effects. A theoretical justification of the application of bootstrap to general semi-parametric M-estimation is provided \cite{ChengHuang10}.  The weighted bootstrap for QR models is considered by \cite{Hahn97}, \citet{BoseChatterjee03}, \cite{ChenPouzo09}, and \cite{BelloniChernozhukovChetverikovFernandezVal19}, the first three in the point-wise case and the later in the uniform context.  For the bootstrap with more complex data, \cite{Hagemann17} develops a wild bootstrap procedure for cluster-robust inference in linear quantile regression models, and \citet{LamarcheParker21} consider a wild residual bootstrap for QR models with panel data.\footnote{For other recent developments in resampling QR, see, e.g., \cite{ChernozhukovFernandezVal05}, \cite{WangvanKeilegomMaidman18}, and \cite{GregoryLahiriNordman18}.} In the time series context, \cite{Fitzenberger98} suggested a moving block bootstrap for inference in QR models.

A second branch of the literature is related to the use of bootstrap methods for standard linear (conditional average) panel data models with FE. \cite{Kapetanios08} discusses bootstrap for panel data when resampling occurs in both cross sectional and time series dimensions, with strictly exogenous regressors and fixed effects, for which the incidental parameter bias does not exist. \cite{Goncalves11} proved the asymptotic validity of the moving blocks bootstrap under general forms of cross sectional and time series dependence in the regression error of a panel linear regression model. More recently, \cite{GoncalvesKaffo15} propose the application of bootstrap methods for inference in linear dynamic panel data models with fixed effects. A treatment of resampling methods when $n$ is large but $T$ is assumed small and fixed can be found in \cite{CameronTrivedi05}. A jackknife technique is applied in \cite{HahnNewey04} for a static nonlinear panel model, and in \cite{DhaeneJochmans15} for a dynamic nonlinear panel model.
This paper extends both of these literatures by establishing the formal properties of random-weighted bootstrap inference for panel data QR models with FE.

The rest of the paper is organized as follows. Section \ref{sec:model} briefly describes the QR model. Section \ref{sec:boot} describes the bootstrap methods and the inference procedures. In Section \ref{sec:asymp} we establish the asymptotic validity of the proposed methods. Section \ref{sec:MC} provides Monte Carlo experiments. In Section \ref{sec:application}, we illustrate empirical usefulness of the new approach by studying the environmental Kuznets curve. Finally, conclusions appear in Section \ref{sec:conclusion}.

\section{Quantile regression with fixed effects}\label{sec:model}

Let $y_{it}$ be a response variable, $\mathbf{x}_{it}$ be a $p$ dimensional vector of explanatory variables, and let $\mathbf{z}_{it}^\top = (1, \mathbf{x}_{it}^\top)$. In this paper we consider the following fixed effects quantile regression (FE-QR) model
\begin{equation}
\label{eq:model}
Q_{\tau}(y_{it} | \mathbf{z}_{it}) = \alpha_{0i}(\tau) + \mathbf{x}_{it}\tr\bm{\beta}_{0}(\tau), \quad i=1,\dots,n, \; t=1,\dots,T,
\end{equation}
where  $Q_{\tau}(y_{it} | \mathbf{z}_{it})$ is the conditional $\tau$-quantile of $y_{it}$ given $\mathbf{z}_{it}$. In model \eqref{eq:model}, $\alpha_{0i}(\tau)$'s are intended to capture some individual-specific source of variability or unobserved heterogeneity that is not adequately controlled for by other explanatory variables. 
We assume that $\tau$ is fixed and for simplicity suppress the dependence on $\tau$ such as $\bm{\beta}_{0}(\tau)= \bm{\beta}_{0}$ and $\alpha_{0i}(\tau) = \alpha_{0i}$.
Throughout the paper, the cross-section dimension is $n$ and the time dimension is $T = T_{n}$ which depends on $n$, but we omit this dependence for simplicity. 

Now we discuss the estimation of model~\eqref{eq:model}. This approach treats each individual effect as a parameter to be estimated.  The FE-QR estimates are those that solve the minimization problem
\begin{equation} \label{feqr_objfun}
(\widehat{\bm{\alpha}},\widehat{\bm{\beta}}) := \argmin_{\bm{\alpha},\bm{\beta}} \frac{1}{nT} \sum_{i=1}^{n} \sum_{t=1}^{T} \rho_{\tau}( y_{it}-\alpha_{i} - \mathbf{x}_{it}\tr\bm{\beta} ),
\end{equation}
where $\rho_{\tau}(u) := u \{ \tau - \1\{u \leq 0\} \}$ is the check function as in \cite{KoenkerBassett78}, $\bm{\beta} \in \R^p$ is the vector of FE-QR slope coefficients and $\bm{\alpha} := (\alpha_{1},...,\alpha_{n})\tr$ is the $n\times1$ vector of individual specific effects or intercepts. This optimization problem can be very large depending on $n$ and $T$.\footnote{It is important to note that in the QR framework there is no general transformation which can eliminate the fixed effects. Thus, we are required to deal with the full problem.}

The use of estimator~\eqref{feqr_objfun} above implies that, unfortunately, even for a linear model of conditional quantiles, the resulting parameter estimates will in general be inconsistent when the number of individuals $n$ goes to infinity while the number of time periods $T$ is fixed. This is a version of the incidental parameters problem \citep{NeymanScott48}. To overcome this drawback, it has become standard in the literature to employ asymptotics for large panels (see, e.g., \cite{Koenker04},  \cite{KatoGalvaoMontes-Rojas12}, and \cite{GalvaoGuVolgushev20}). In particular, it is common to derive the asymptotic properties of the estimator and associated test statistics under an assumption that $n$ and $T$ grow simultaneously.

The asymptotic properties of the FE-QR estimator have been established in the literature, for instance in \cite{KatoGalvaoMontes-Rojas12} and \cite{GalvaoGuVolgushev20}. They show consistency and asymptotic normality of the estimator under a sample size growth condition.
In particular,  suppose that the observations $(y_{it}, \mathbf{z}_{it})$ are independent across units indexed by $i$ but may be serially dependent (generally stationary and $\beta$-mixing).  Let
\begin{equation}
  f_i(y | \mathbf{z}) := f_{y_{i1} | \mathbf{z}_{i1}}(y + Q_{y_{i1} | \mathbf{z}_{i1}}(\tau | \mathbf{z}) | \mathbf{z})
\end{equation}
denote the conditional density of $y_{i1} - Q_{y_{i1}|\mathbf{z}_{i1}}(\tau | \cdot )$ given $\mathbf{z}$, and let $f_i(y) := \E [ f_i(y | \mathbf{z}_{i1}) ]$ be its marginal density function.  Further define
\begin{align}
  \mathbf{g}_i &:= \E[f_i (0 | \mathbf{z}_{i1}) \mathbf{x}_{i1}] / f_i(0), \label{eq:def-gi} \\
  \Gamma_n &:= \frac{1}{n}\sum_{i=1}^n \E[f_i (0 | \mathbf{z}_{i1}) \mathbf{x}_{i1}(\mathbf{x}_{i1}-\mathbf{g}_i)^\top], \label{eq:def-gamman} \\
  V_{n} &:= \frac{1}{n}\sum_{i=1}^n \Var\Big(T^{-1/2}\sum_{t=1}^T (\mathbf{x}_{it}- \mathbf{g}_i)(\tau - \1\{y_{it} \leq \alpha_{i0} + \mathbf{x}_{it}^\top \bm{\beta}_{0\tau}\})\Big). \label{eq:def-Vn}
\end{align}
Assume that $\Gamma_{n}$ is nonsingular for each $n$ and the limits $\Gamma := \lim_{n \to \infty}\Gamma_{n}$ and $V:= \lim_{n \to \infty} V_{n}$ exist and are nonsingular.  Then under mild regularity conditions and the rate restriction $n(\log T)^{4}/T \to 0$, \cite{GalvaoGuVolgushev20} show that
\begin{equation*}
\sqrt{nT}\left( \widehat{\bm{\beta}} - \bm{\beta}_{0} \right) =\Gamma _{n}^{-1}\frac{1}{\sqrt{nT}}\sum_{i=1}^{n} \sum_{t=1}^{T} (\mathbf{x}_{it}-\mathbf{g}_i) \psi _{\tau} \left( e_{it} \right) + o_{p}(1),
\end{equation*}
where $\psi_{\tau}(u) := \{ \tau - \1\{u \leq 0\} \} $ and $e_{it} = y_{it} - \alpha_{i0} - \mathbf{x}_{it}^\top \bm{\beta}_0$. Moreover, $\sqrt{nT} \left( \widehat{\bm{\beta}} - \bm{\beta}_{0} \right) \stackrel{d}{\to} N  \left( \bm{0}, \Sigma \right)$, where
\begin{equation}
  \Sigma:=\Gamma^{-1} V \Gamma^{-1}.
\end{equation}
Notice that each component of the variance-covariance matrix is a function of the quantile level $\tau$, but we suppress this dependence for simplicity.

As noted above, one crucial condition in \cite{GalvaoGuVolgushev20} is that the time dimension $T$ grows slightly faster than $n$. This implies that the estimator is expected to work well in applications where $T$ is slightly larger than $n$. This condition on $T$ for asymptotic normality of the FE-QR estimator is very similar to the standard one (i.e., $n/T \to 0$) that is found in the nonlinear panel data literature (see, e.g., \cite{HahnNewey04}).

Alternative options have been considered for constructing confidence intervals and conducting inference for the common parameters of interest $\bm{\beta}_{0}$. Most straightforwardly, one can compute an estimate of the asymptotic variance-covariance matrix $\Sigma$ and construct confidence intervals directly from it. However, the asymptotic covariance matrix of the panel QR estimator given in the literature (see, e.g.,  \cite{GalvaoKato17}) depends on the conditional density of the error term, which requires selecting a bandwidth, and it may not be easy to estimate. We propose to use a bootstrap procedure to facilitate statistical inference for the parameters of interest in the FE-QR model. The next section describes the proposed random-weighted bootstrap method for inference procedures. 

\section{Bootstrap}\label{sec:boot}

The main concern of this section is the application of weighted bootstrap procedures to the problem of constructing confidence intervals and conducting inference for the common parameters, $\bm{\beta}_{0}$, of the FE-QR model in equation \eqref{eq:model}, based on the estimation methods described in the previous section.

Traditionally, bootstrap techniques have been successfully employed to construct confidence intervals for QR in the cross-section context.  In this paper we use a random-weighted bootstrap commonly used for cross-sectional resampling. This scheme consists in resampling weights from the cross-section dimension, maintaining intact the temporal structure for each individual $i$.

\subsection{A random-weighted bootstrap method}

We consider the following random-weighted bootstrap method to approximate the distribution of the panel QR estimator.

Let $\{\omega _{i},i=1,\cdot \cdot \cdot ,n\}$ be $i.i.d.$ non-negative random weights with mean and variance both equal to one. For each draw of $\{\omega _{i},i=1,\cdot \cdot \cdot ,n\}$, we estimate the bootstrap coefficients $(\widehat{\bm{\alpha}}^{\ast },\widehat{\bm{\beta}}^{\ast })$ using a weighted FE-QR objective:
\begin{equation}\label{eq:BQRFE}
  (\widehat{\bm{\alpha}}^{\ast},\widehat{\bm{\beta}}^{\ast})= \argmin_{\bm{\alpha}, \bm{\beta} }\sum_{i=1}^{n}\omega _{i}\sum_{t=1}^{T} \rho _{\tau }(y_{it}-\alpha _{i}-\mathbf{x}_{it}\tr\bm{\beta} ).
\end{equation}
After conducting this procedure over many repetitions, we approximate the distribution of $(\widehat{\bm{\beta}}-\bm{\beta}_0)$ by that of $(\widehat{\bm{\beta}}^{\ast}- \widehat{\bm{\beta}})$.

It should be noted that the usual pairwise bootstrap -- sampling over $(i,t)$ from $\{ (y_{it},\mathbf{x}_{it}) \}$ with replacement --  would only be appropriate when the data are $i.i.d.$ Since $\{ (y_{it},\mathbf{x}_{it}) \}$ is not $i.i.d.$ because of the individual effects and potential serial correlation, the usual pairwise bootstrap is not directly applicable to panel data. However, the random-weighted method is applicable to a wide variety of models and not limited to $i.i.d.$ sampling.

\subsection{Practical implementation}
The practical implementation of the random-weighted bootstrap method is very simple. The main algorithm for implementing the methods is as following. Take $B$ as a large integer:
\begin{itemize}
\item[] Step 1. Draw $i.i.d.$ non-negative random weights $\{\omega_{i}, i=1,...,n \}$ from the appropriate distribution satisfying mean and variance both equal to one.

\item[] Step 2. For a given quantile of interest $\tau\in(0,1)$, fit the weighted FE-QR panel model in equation \eqref{eq:BQRFE} using the sample, $(y_{it},\mathbf{x}_{it})$ and weights $\omega_{i}$. Denote the bootstrap estimate by $(\widehat{\bm{\alpha}}^{\ast},\widehat{\bm{\beta}}^{\ast})$.

\item[] Step 3. Repeat Steps 1--2 $B$ times.

\item[] Step 4. Approximate the distribution of $\sqrt{nT} (\widehat{\bm{\beta}} - \bm{\beta}_{0})$ by the empirical distribution of the $B$ observations of $\sqrt{nT}(\widehat{\bm{\beta}}^{\ast} - \widehat{\bm{\beta}})$.

\end{itemize}

\subsubsection*{Percentile confidence interval}

Note that by choosing the number of bootstrap simulations $B$ in the algorithm above large enough, the distribution of $\widehat{\bm{\beta}}^{\ast} - \widehat{\bm{\beta}}$ can be computed with any desired precision.  The distribution function of $\widehat{\bm{\beta}}^\ast - \widehat{\bm{\beta}}$ can be used to estimate the distribution function of $\widehat{\bm{\beta}} - \bm{\beta}_0$.  Specifically, suppose that the parameter of interest $\beta$ is scalar and let $\widehat{G}$ be the cumulative distribution function of $\widehat{\beta}^{\ast} - \widehat{\beta}$. Then, for a given quantile level $\tau$, we compute the $1-\lambda$ percentile confidence interval for each element in the coefficient vector $\beta_{0}(\tau)$ by the $\lambda/2$ and $1-\lambda/2$ sample percentiles of $\widehat{G}$:
\begin{equation} \label{eq:boot_percentile}
CI_P = %[\widehat{\beta}^{\ast}_{\lambda/2}, \widehat{\beta}^{\ast}_{1-\lambda/2}]_p =
 [\widehat{G}^{-1}(\lambda / 2), \widehat{G}^{-1}(1-\lambda / 2)].
\end{equation}
These percentiles may be used as to estimate the endpoints of a confidence interval for $\bm{\beta}_0$.\footnote{In Monte Carlo simulations, we compared this method with percentile CIs of the form $(2\widehat{\bm{\beta}} - \widehat{G}^{-1}(1-\lambda / 2), 2\widehat{\bm{\beta}} - \widehat{G}^{-1}(\lambda / 2))$, but that performed poorly compared to the interval described in~\eqref{eq:boot_percentile}.}

\subsubsection*{Variance-covariance matrix estimation}

For a fixed quantile level $\tau$, we define the bootstrap estimate of the asymptotic covariance matrix $\Sigma$ given bootstrap realizations $\{\widehat{\bm{\beta}}^{\ast b}\}_{b=1}^B$ as
\begin{equation}\label{eq:boot_var}
\widehat{\Sigma}^{\ast} = \frac{1}{B} \sum_{b=1}^{B} (\widehat{\bm{\beta}}^{\ast b} - \widehat{\bm{\beta}})(\widehat{\bm{\beta}}^{\ast b} - \widehat{\bm{\beta}})^{\top}.
\end{equation}
The standard errors of $\widehat{\bm{\beta}}$ are the square roots of the diagonal elements of $\widehat{\Sigma}^{\ast}$. Given this estimated covariance matrix, testing general hypotheses $R\bm{\beta}_0=r$ for the vector $\bm{\beta}_0$ can be accommodated by Wald-type tests.

There are two options for the construction of confidence intervals given consistent bootstrap covariance matrix estimates.  Once again, assume that $\beta$ is scalar for simplicity.  Let $se^\ast$ be 
$\sqrt{\widehat{\Sigma}^{\ast}}$ (in higher dimensions it is the square root of a diagonal element). We could use the estimated standard errors directly in the confidence interval as following
\begin{equation}\label{eq:boot_se}
  CI_{SE} = [\widehat{\beta} - z_{1 - \lambda/2} se^\ast, \widehat{\beta} - z_{1 - \lambda/2} se^\ast],
\end{equation}
where $z_{\lambda}$ denotes the $\lambda$-th quantile of the standard normal distribution.  Alternatively, we may compute a reference distribution of bootstrap $t$ statistics where $t_b^\ast = (\widehat{\beta}^\ast - \widehat{\beta}) / se^\ast$, and construct the following confidence interval
\begin{equation}\label{eq:boot_ref_dist}
  CI_{RD} = [\widehat{\beta} - t^\ast_{1 - \lambda/2} \widehat{se}, \widehat{\beta} - t^\ast_{1 - \lambda/2} \widehat{se}],
\end{equation}
where similarly, $t^\ast_\lambda$ denotes the (empirical) $\lambda$-th quantile of the bootstrap $t$ distribution, and $\widehat{se}$ is the $\sqrt{\widehat{\Sigma}}$ with $\widehat{\Sigma}$ being the variance of $\widehat{\beta}$ estimated from the original sample (in higher dimensions it is the square root of a diagonal element).  This interval suffers from the defect that we would need to estimate the standard error of $\widehat{\beta}$, which was we were hoping to avoid by using the bootstrap in this setting.  However, confidence intervals of this form are included in the simulation study for comparison so we describe them here.

In the next section we formally establish the asymptotic properties of the proposed random-weighted bootstrap method and its validity for inference procedures.

%\begin{remark}
%\textcolor{blue}{(POTENTIAL REMARK ON UNBALANCED PANEL)}
%\end{remark}

\section{Asymptotic validity of the bootstrap}\label{sec:asymp}

After discussing basic regularity conditions, we discuss two main results.  First we discuss the consistency of the bootstrap estimate of the distribution function of $\widehat{\bm{\beta}} - \bm{\beta}_0$, and next, the consistency of the bootstrap estimator of the covariance of $\widehat{\bm{\beta}} - \bm{\beta}_0$.  Consistency of the distribution estimator does not imply that of the second moment estimator, and we discuss the way in which conditions must be strengthened to maintain second-moment estimator consistency.

\subsection{Basic assumptions}\label{sec:ass}

% Throughout the paper, we use the following notations: for a square matrix $A$, let $\|A\|$ denote the maximal absolute eigenvalue of $A$, while for a vector $v$, $\|v\|$ denotes the usual Euclidean norm. 

We use $\P^\ast$ and $\E^\ast$ to denote the probability measure and expected value take with respect to the bootstrapped data conditional on the observations, and use $X_n^\ast \pconvs X$ to denote convergence in probability of a sequence of bootstrap statistics to a limit conditional on the observations.  Recall that $\mathbf{z}_{it}^\top = (1, \mathbf{x}_{it}^\top)$ is a vector of dimension $p+1$ and let $\mathcal{Z}$ denote the support of $\mathbf{z}_{it}$. We make the following assumptions.

%\begin{itemize}
\begin{enumerate}[label=(A\arabic*), ref=(A\arabic*)]
  \setcounter{enumi}{-1}
  \item \label{a:data} For each $i \geq 1$, the process $\{ (y_{it},\mathbf{x}_{it}) : t \in \mathbb{Z} \}$ is strictly stationary and $\beta$-mixing. Let $\beta_{i}(j)$ denote the $\beta$-mixing coefficient of the process $\{ (y_{it},\mathbf{x}_{it}) :t \in \mathbb{Z}  \}$. Assume that there exist constants $b_\beta \in (0,1), C_\beta > 0$ independent of $i$ such that $\sup_i \beta_i(j) \leq C_{\beta} b_\beta^j =: \beta(j) \quad \forall j \geq 1.$
  \item \label{a:boundedZ} Assume that $\|\mathbf{z}_{it}\| \leq M < \infty$ almost surely, that $c_{\lambda}\leq\lambda_{\min}(\E [\mathbf{z}_{it} \mathbf{z}_{it}^\top])\leq\lambda_{\max}(\E [\mathbf{z}_{it} \mathbf{z}_{it}^\top])\leq C_{\lambda}$ holds uniformly in  $i$ for some fixed constants $c_{\lambda}>0$ and $C_{\lambda} <\infty$ and that $(\alpha_i, \bm{\beta})$ lies in a compact set for all $i$.
  \item \label{a:boundedf} The conditional distribution $F_{y_{i1}|\mathbf{z}_{i1}}(y|\mathbf{z})$ is twice differentiable w.r.t. $y$, with the corresponding derivatives $f_{y_{i1}|\mathbf{z}_{i1}}(y|\mathbf{z})$ and $f'_{y_{i1}|\mathbf{z}_{i1}}(y|\mathbf{z})$. Assume that 
    \begin{equation*}
      f_{\text{max}} :=\sup_i \sup_{y \in \mathbb{R},\mathbf{z}\in \mathcal{Z}} |f_{y_{i1}|\mathbf{z}_{i1}}(y|\mathbf{z})| < \infty,
    \end{equation*}
    and 
    \begin{equation*}
      \overline{f'} := \sup_i \sup_{y \in \mathbb{R},\mathbf{z}\in \mathcal{Z}} |f'_{y_{i1}|\mathbf{z}_{i1}}(y|\mathbf{z})| < \infty.
    \end{equation*}
  \item \label{a:minf} Denote by $\mathcal{T}$ an open neighborhood of $\tau$. Assume that uniformly across $i$, there exists a constant $f_{\min} < f_{\max}$ such that
    \begin{equation*}
      0 < f_{\min} \leq \inf_i \inf_{\eta \in \mathcal{T}} \inf_{\mathbf{z} \in \mathcal{Z}} f_{y_{i1}|\mathbf{z}_{i1}}(Q_{y_{i1} | \mathbf{z}_{i1}}(\eta | \mathbf{z}) | \mathbf{z}).
    \end{equation*}
\end{enumerate}
\begin{enumerate}[label=(B\arabic*), ref=(B\arabic*)]
  \item \label{a:ydens} For each $i=1,...,n$ and $j>1$, the random vector $(y_{i1},y_{i (1+j)})$ has a density conditional on $(\mathbf{z}_{i1},\mathbf{z}_{i (1+j)})$ and this density is bounded uniformly across $i,j$.
\end{enumerate}

\begin{enumerate}[label=(C\arabic*), ref=(C\arabic*)]
\item \label{a:bweight} $\{\omega _{i},i=1,\cdot \cdot \cdot ,n\}$ are $i.i.d.$ positive random weights with mean and variance both equal to one.
\end{enumerate}

Condition~\ref{a:data} assumes that the data are independent across individuals, and strictly stationary within each individual. It allows for stationary $\beta$-mixing which is used in \cite{KatoGalvaoMontes-Rojas12, GalvaoWang15} and is similar to \cite{HahnKuersteiner11}. %The $\beta$-mixing condition is stronger than $\alpha$-mixing. Nevertheless, $\beta$-mixing is still satisfied in a reasonably large class of time series models.\footnote{For example, consider the MA $(\infty)$ process $X_{t}= \sum_{j=0}^{ \infty}a_{j} \varepsilon_{t-j}$, where $a_{j} \to 0$ exponentially fast (note that the ARMA($p,q$) process, subject to standard assumptions, fulfils this condition), and $\{ \varepsilon_{t} \}$ is $i.i.d.$ If the density function of $\varepsilon_{t}$ exists, then $\{ X_{t}\}$ is $\beta$-mixing with $\beta(n) \to 0$ exponentially fast.}

Condition~\ref{a:boundedZ} poses a boundedness condition on the norm of the regressors, which is also standard in the literature, see for instance \cite{Koenker04, KatoGalvaoMontes-Rojas12, GalvaoWang15}. Condition~ \ref{a:boundedZ} also assures that the eigenvalues of $\E [\mathbf{z}_{it} \mathbf{z}_{it}^\top]$ are bounded away from zero and infinity uniformly across $i$.  Similar assumptions were made in \cite{ChaoVolgushevCheng17}.
Conditions~\ref{a:boundedf} and~\ref{a:minf} impose smoothness and boundedness of the conditional distribution, the density and its derivatives. The same type of assumption has been imposed in \cite{GalvaoWang15}. \cite{ChaoVolgushevCheng17} also make similar assumptions when deriving Bahadur representations for QR estimators in a setting without panel data.  Condition~\ref{a:ydens} is needed because the data are not $i.i.d.$ and we need to impose a condition on the joint distributions; similar conditions were imposed in \cite{KatoGalvaoMontes-Rojas12, GalvaoWang15}.  
Finally, assumption~\ref{a:bweight} is a common bootstrap weight condition  (see, e.g., \cite{BelloniChernozhukovChetverikovFernandezVal19} and \cite{RaoZhao92}).

To state the asymptotic properties of $\widehat{\bm{\beta}}^{\ast}$ we make the following additional assumption.

\begin{enumerate}[label=(FD), ref=(FD)]
  \item \label{a:Avar} Assume that $\Gamma_n$ in equation \eqref{eq:def-gamman}  is non-singular for each $n$ and that $\Gamma :=\lim_{n \to \infty} \Gamma_n$ exists and is non-singular. Further assume that $V := \lim_{n \to \infty} V_{n}$ exists and is non-singular, where $V_{n}$ is defined in equation \eqref{eq:def-Vn}.
\end{enumerate}

This assumption was also made by \cite{KatoGalvaoMontes-Rojas12} (see their condition (D3)), it involves the long run covariance matrix of the leading piece in the Bahadur representation for $\widehat{\bm{\beta}}$.

Finally, we assume that the observations and bootstrap weights satisfy slightly stronger moment conditions for consistent covariance matrix estimation.

\begin{enumerate}[label=(CV), ref=(CV)]
  \item \label{a:moments} Assume that for some $\epsilon > 0$, $\E [ \| \mathbf{x}_{it} \|^{2 + \epsilon} ] < \infty$ and $\E [ \omega_i^{2 + \epsilon} ] < \infty$.
\end{enumerate}

    This assumption is made to ensure uniform square integrability of the sequence of bootstrap variance estimates, which is only slightly stronger than assumption~\ref{a:Avar} above and ensures the convergence of bootstrap moments to finite limits (this is discussed more in the remark below Theorem~\ref{thm:consistent_var} below).

\subsection{Asymptotic results}\label{sec:asymp_results}

    Now we formally establish the asymptotic validity of the proposed random-weighted bootstrap method. The following result show consistency of the distribution of the bootstrap estimate.

\begin{theorem}[Consistency of the Bootstrap] \label{thm:consistent} Assume that $n (\log T)^{4}/T \to 0$ as $n \to \infty$, and $T$ grows at most polynomially in $n$. Under conditions \ref{a:data}-\ref{a:minf}, \ref{a:ydens}, \ref{a:bweight}, and \ref{a:Avar}, for fixed $\tau$, it follows that 
\begin{equation*}
\sup_{x} \left| \P^{\ast} \left( \sqrt{nT}(\widehat{\bm{\beta}}^{\ast} - \widehat{\bm{\beta}}) \leq x \right) - \P\left(\sqrt{nT} (\widehat{\bm{\beta}} - \bm{\beta}_{0}) \leq x \right) \right| \pconv 0.
\end{equation*}
\end{theorem}

The consistency of a bootstrap confidence interval is closely related to the consistency of the bootstrap estimator of the distribution of $\sqrt{nT}(\widehat{\bm{\beta}} -\bm{\beta}_{0})$. Theorem \ref{thm:consistent} states that the bootstrap estimator for the distribution is consistent relative to the Kolmogorov-Smirnov distance. Consistency relative to the Kolmogorov-Smirnov distance is equivalent to the requirement that, uniformly in $x$,
\begin{equation*}
  \P\left(\sqrt{nT}(\widehat{\bm{\beta}} - \bm{\beta}_{0}) \leq x \right) \rightarrow F(x), \quad \P^{\ast}\left(\sqrt{nT}(\widehat{\bm{\beta}}^{\ast} - \widehat{\bm{\beta}}) \leq x \right) \rightarrow F(x).
\end{equation*}
It has been shown in the literature~--- see, e.g., \citet{GalvaoGuVolgushev20}~--- that $\sqrt{nT}(\widehat{\bm{\beta}} -\bm{\beta}_{0})$ converges in distribution to a Gaussian distribution, implying that $F$ is actually the CDF of a normal distribution with variance $\Sigma$.

This distributional consistency implies the consistency of percentile confidence intervals in equation \eqref{eq:boot_percentile} above. Although Theorem~\ref{thm:consistent} implies consistency of the bootstrap percentile confidence intervals, it does not necessarily imply consistency of variance-covariance estimation \citep{GhoshParrSinghBabu84}.  Theorem~\ref{thm:consistent_var} implies that the variance can be consistently estimated using the estimator shown in equation~\eqref{eq:boot_var}.  The result in Theorem \ref{thm:consistent_var} is very useful in empirical applications since it allows one to easily estimate the variance-covariance matrix by resampling procedures without calculating each of its components separately.
\begin{theorem}[Consistency of the variance-covariance matrix] \label{thm:consistent_var}
  Under conditions of Theorem \ref{thm:consistent} and Assumption~\ref{a:moments}, for fixed $\tau$,
\begin{equation*}
  \widehat{\Sigma}^{\ast} \pconvs \Sigma.
\end{equation*}
\end{theorem}

\begin{remark}
  The result of Theorem~\ref{thm:consistent_var} is slightly stronger than that of Theorem~\ref{thm:consistent}.  The weak convergence (conditional on the observations) of $\sqrt{nT}(\widehat{\bm{\beta}} - \widehat{\bm{\beta}})$ to a limit does not imply that the sequence of bootstrap second moments converges as well.  One way to see why is to consider a Bahadur representation that is (indirectly) derived in the proof of Theorem~\ref{thm:consistent}:
  \begin{equation}
    \sqrt{nT} \left( \widehat{\bm{\beta}}^\ast - \widehat{\bm{\beta}} \right) = (\widehat{\Gamma}_n^\ast)^{-1} \sum_{i=1}^n \omega_i \sum_{t=1}^T (\mathbf{x}_{it} - \bar{\mathbf{g}}_{Ti}) \psi_\tau(\widehat{e}_{it}) + o_{p^\ast}(1).
  \end{equation}
  This expression depends on a few terms that are defined in the proof of Theorem~\ref{thm:consistent} but are inessential here.  The conditions used to show distributional consistency must be strengthened for variance estimation to ensure that the covariance matrix may be consistently estimated.  First, the terms in the sum must be uniformly square integrable so that this main part of the Bahadur representation converges.  Second, we must also ensure that the remainder term is square integrable, which is not implied by the form of asymptotic negligibility implied by its $o_{p^\ast}(1)$ characterization~--- although the $o_{p^\ast}(1)$ characterization implies that the remainder is small with high probability, it would be a problem if with low probability, the remainder were extremely large.  The moment conditions specified in Assumption~\ref{a:moments} are sufficient to ensure that both of these requirements are satisfied.
\end{remark}

\section{Monte Carlo}\label{sec:MC}

In this section, we report a simulation study to assess the finite sample performance of the proposed bootstrap inference procedures.

\subsection{Designs}

We consider several different simulation designs. For static models we generate data from a simple version of model \eqref{eq:model}:
\begin{equation}\label{eq:mc1}
y_{it} = \alpha_{i} + \beta x_{it} + (1+\gamma x_{it}) \epsilon_{it}, \quad i = 1,\dots,n, \; t= 1,\dots,T.
\end{equation}
In all designs, we set $\alpha_{i} \sim \text{i.i.d.} \ U[0, 1]$, $x_{it} = 0.3 \alpha_{i} + z_{it}$ with $z_{it} \sim \text{i.i.d.} \chi_{3}^{2}$ and $\beta=1$.  We use $\gamma \in \{0, 0.2\}$ to generate data following location and location-scale models respectively.  We generate static models using independent or serially dependent error terms.  For i.i.d. error designs we simulate $\epsilon_{it} \sim \text{i.i.d.} \chi_{4}^{2}$, and in models with serial dependence, we use an ARMA$(1, 1)$ model in the innovation term: $\epsilon_{it} = \rho \epsilon_{it-1} + \varepsilon_{it} + \theta \varepsilon_{it-1}$. We set $\rho = 0.4$ and $\theta = 0.5$ when generating dependent errors. Note that ARMA variables with continuous distribution are geometrically $\beta$-mixing \citep{Mokkadem88}.

To examine how the estimators behave in a dynamic model, we consider the autoregressive model
\begin{equation}
\label{eq:mc3}
y_{it} = \alpha_{i} + \rho y_{it-1} +  \epsilon_{it},
\end{equation}
where the innovations $\epsilon_{it} \sim \text{i.i.d.} \chi_{4}^{2}$.  For these simulations we set $\rho = 0.4$, and in generating $y_{it}$ we set $y_{i,-50}=0$ and discard the first 50 observations, using the observations $t = 0$ through $T$ for estimation.  

In the stationary model in \eqref{eq:mc1} let $Q_\epsilon(\tau)$ denote the $\tau$th quantile of the stationary distribution of $\epsilon$, the $\alpha_{i0}(\tau) = \alpha_{i} + Q_{\epsilon}^{-1}(\tau)$ and $\beta_{0}(\tau)= \beta + \gamma Q_{\epsilon}(\tau)$.  Then $Q_\epsilon(\tau) = F_\epsilon^{-1}(\tau)$, the $\tau$th quantile of the marginal distribution of $\epsilon_{it}$.  In the dynamic model in \eqref{eq:mc3} the true quantile regression coefficients at the $\tau$th quantile are $\alpha_{i0}(\tau) = \alpha_{i} + F^{-1}_\epsilon(\tau)$ and $\beta_{0}(\tau) = \rho$. 

For all experiments, we consider sample sizes $n \in \{ 25, 50, 100 \}$ and $T \in \{ 20, 50, 100 \}$. We estimate the model for the three quartiles $\tau\in\{0.25,0.50,0.75\}$.  We compute the random-weighted bootstrap procedure discussed in Section \ref{sec:boot} for each experiment. The random weights follow an exponential distribution, $\omega_{i} \sim \text{exp}(1)$ and the number of bootstrap repetitions is always 999.\footnote{We performed a simulation with bootstrap repetitions that increased with the sample size and found nearly the same results while taking much longer because of the large number of simulation designs and repetitions.  In actual data analysis, it is best to allow the number of simulation repetitions to be as large as is practical.}  The results in the tables are based on 1,000 simulation replications.

In all the simulations, the performance of the confidence intervals is evaluated with the empirical coverage, that is, the proportion of simulations in which the confidence interval (CI) contains the true parameter. Throughout the simulation study, the nominal coverage probability for for all CIs is $90\%$.
%and the root mean squared error (RMSE) of the estimator of the variance of $\widehat{\beta}$.

\VerbatimFootnotes % used by the fancyvrb package.
We report results for a few different confidence intervals for $\beta_0$. First, coverage for the percentile interval $CI_P$ defined in~\eqref{eq:boot_percentile} is denoted {\bf RWB.p}. Second, coverage for CIs using the bootstrap variance-covariance matrix, $CI_{SE}$ described in~\eqref{eq:boot_se} are denoted {\bf RWB.se}.  We used the Powell estimate of the standard error to implement these confidence intervals.  Third, we report bootstrap results to estimate a reference distribution using the empirical quantile of the bootstrapped $t$ statistics, the $CI_{RD}$ described in~\eqref{eq:boot_ref_dist}, and denote those by {\bf RWB.t}. Here the bootstrap $t$ distribution also uses a Powell estimate of the standard error with each bootstrap sample. Finally, we report results for the CI using standard asymptotic theory to estimate the CI using a Powell estimate for covariance estimation, labeling these by {\bf AT}.\footnote{For estimation of the asymptotic covariance matrix, we use the Gaussian kernel and the default bandwidth option that one would use by default in the {\tt quantreg} package in \texttt{R} by choosing the standard error option \verb@se = "ker"@.}

\subsection{Results}

\subsubsection{Independent case}

The results for the independent case are reported in Tables \ref{chisq_ind_loc_cover}, and \ref{chisq_ind_het_cover}, for the location shift and location-scale shift cases, respectively.

The first important result in Table \ref{chisq_ind_loc_cover}, regarding the location case, is that the empirical coverage for the percentile bootstrap \textbf{RWB.p} approximates the nominal coverage very well. The results is approximately the same for the \textbf{RWB.se}. Moreover, the table shows that for these two cases the results are very similar for all the three quantiles considered. Another important result from Table \ref{chisq_ind_loc_cover} is that both \textbf{RWB.p} and \textbf{RWB.se} produce empirical coverage close to the nominal for relatively small samples, that is, when $n$ is small relative to $T$. But, as expected from the theory, the results improve as the sample size increases. 

Table \ref{chisq_ind_loc_cover} also indicates some distortions in the empirical coverage for \textbf{RWB.t}, which is especially severe when considering $\tau=3/4$ and cases $T=20$ with $n=25$ and $n=50$.

%latex.default(cover_tab, title = "Coverage prob.", file = paste0("./tables/",     file_prefix, "_cover.tex"), label = paste0(file_prefix, "_cover"),     size = "scriptsize", cgroup = c("", method_names), n.cgroup = c(2,         rep(ntaus, (ncol(cover_tab) - 2)/ntaus)), n.rgroup = rep(nT,         nn), caption = paste(desc, "Nominal 90 percent CIs.  1000 simulation\n                  repetitions used.  RWB.p, RWB.se and RWB.t are randomly\n                  weighted bootstrap confidence intervals using percentile,\n                  bootstrap standard error estimate or bootstrap t estimates\n                  respectively, and AT stands for asymptotic theory."))%
\begin{table}[!tbp]
{\scriptsize
\caption{Homoskedastic, independent error. Nominal 90 percent CIs.  1000 simulation
                  repetitions used.  RWB.p, RWB.se and RWB.t are randomly
                  weighted bootstrap confidence intervals using percentile,
                  bootstrap standard error estimate or bootstrap t estimates
                  respectively, and AT stands for asymptotic theory.\label{chisq_ind_loc_cover}} 
\begin{center}
\begin{tabular}{rrcrrrcrrrcrrrcrrr}
\hline\hline
\multicolumn{2}{c}{\bfseries }&\multicolumn{1}{c}{\bfseries }&\multicolumn{3}{c}{\bfseries RWB.p}&\multicolumn{1}{c}{\bfseries }&\multicolumn{3}{c}{\bfseries RWB.se}&\multicolumn{1}{c}{\bfseries }&\multicolumn{3}{c}{\bfseries RWB.t}&\multicolumn{1}{c}{\bfseries }&\multicolumn{3}{c}{\bfseries AT}\tabularnewline
\cline{4-6} \cline{8-10} \cline{12-14} \cline{16-18}
\multicolumn{1}{c}{n}&\multicolumn{1}{c}{T}&\multicolumn{1}{c}{}&\multicolumn{1}{c}{1/4}&\multicolumn{1}{c}{1/2}&\multicolumn{1}{c}{3/4}&\multicolumn{1}{c}{}&\multicolumn{1}{c}{1/4}&\multicolumn{1}{c}{1/2}&\multicolumn{1}{c}{3/4}&\multicolumn{1}{c}{}&\multicolumn{1}{c}{1/4}&\multicolumn{1}{c}{1/2}&\multicolumn{1}{c}{3/4}&\multicolumn{1}{c}{}&\multicolumn{1}{c}{1/4}&\multicolumn{1}{c}{1/2}&\multicolumn{1}{c}{3/4}\tabularnewline
\hline
&&&&&&&&&&&&&&&&&\tabularnewline
$ 25$&$ 20$&&$86.4$&$88.8$&$88.1$&&$89.7$&$88.7$&$88.1$&&$88.7$&$84.6$&$74.4$&&$97.6$&$94.3$&$85.0$\tabularnewline
$ 25$&$ 50$&&$86.2$&$85.5$&$87.8$&&$87.9$&$86.6$&$87.3$&&$89.5$&$83.0$&$83.3$&&$95.6$&$91.1$&$89.0$\tabularnewline
$ 25$&$100$&&$85.1$&$86.8$&$84.3$&&$86.0$&$85.8$&$86.3$&&$87.4$&$85.2$&$81.8$&&$93.1$&$89.9$&$86.8$\tabularnewline
\hline
&&&&&&&&&&&&&&&&&\tabularnewline
$ 50$&$ 20$&&$88.7$&$87.3$&$88.1$&&$87.4$&$88.3$&$88.3$&&$88.0$&$83.9$&$79.9$&&$96.2$&$91.2$&$86.2$\tabularnewline
$ 50$&$ 50$&&$87.8$&$86.1$&$87.1$&&$87.1$&$87.1$&$86.6$&&$89.5$&$83.4$&$84.9$&&$93.4$&$88.7$&$87.0$\tabularnewline
$ 50$&$100$&&$88.8$&$89.1$&$89.6$&&$88.6$&$88.6$&$89.5$&&$87.8$&$87.5$&$86.8$&&$92.7$&$90.3$&$91.1$\tabularnewline
\hline
&&&&&&&&&&&&&&&&&\tabularnewline
$100$&$ 20$&&$87.7$&$88.7$&$88.3$&&$88.8$&$87.9$&$88.1$&&$87.8$&$83.6$&$81.8$&&$94.9$&$89.6$&$86.8$\tabularnewline
$100$&$ 50$&&$88.5$&$89.2$&$89.7$&&$88.3$&$89.3$&$88.1$&&$88.7$&$86.0$&$85.8$&&$91.8$&$90.5$&$89.0$\tabularnewline
$100$&$100$&&$88.1$&$89.4$&$90.1$&&$88.1$&$88.1$&$89.5$&&$88.0$&$86.1$&$87.8$&&$90.9$&$89.3$&$90.8$\tabularnewline
\hline
\end{tabular}\end{center}}
\end{table}

Table \ref{chisq_ind_het_cover} collects the results for the location-scale case. The overall patterns are very similar to those in Table \ref{chisq_ind_loc_cover}, with a very slight reduction in empirical coverages. Both \textbf{RWB.p} and \textbf{RWB.se} procedures produce consistent results across quantiles, empirical coverages are close to nominal 90\% for small samples, and coverage improves as sample size increases. 

%latex.default(cover_tab, title = "Coverage prob.", file = paste0("./tables/",     file_prefix, "_cover.tex"), label = paste0(file_prefix, "_cover"),     size = "scriptsize", cgroup = c("", method_names), n.cgroup = c(2,         rep(ntaus, (ncol(cover_tab) - 2)/ntaus)), n.rgroup = rep(nT,         nn), caption = paste(desc, "Nominal 90 percent CIs.  1000 simulation\n                  repetitions used.  RWB.p, RWB.se and RWB.t are randomly\n                  weighted bootstrap confidence intervals using percentile,\n                  bootstrap standard error estimate or bootstrap t estimates\n                  respectively, and AT stands for asymptotic theory."))%
\begin{table}[!tbp]
{\scriptsize
\caption{Heteroskedastic, independent error. Nominal 90 percent CIs.  1000 simulation
                  repetitions used.  RWB.p, RWB.se and RWB.t are randomly
                  weighted bootstrap confidence intervals using percentile,
                  bootstrap standard error estimate or bootstrap t estimates
                  respectively, and AT stands for asymptotic theory.\label{chisq_ind_het_cover}} 
\begin{center}
\begin{tabular}{rrcrrrcrrrcrrrcrrr}
\hline\hline
\multicolumn{2}{c}{\bfseries }&\multicolumn{1}{c}{\bfseries }&\multicolumn{3}{c}{\bfseries RWB.p}&\multicolumn{1}{c}{\bfseries }&\multicolumn{3}{c}{\bfseries RWB.se}&\multicolumn{1}{c}{\bfseries }&\multicolumn{3}{c}{\bfseries RWB.t}&\multicolumn{1}{c}{\bfseries }&\multicolumn{3}{c}{\bfseries AT}\tabularnewline
\cline{4-6} \cline{8-10} \cline{12-14} \cline{16-18}
\multicolumn{1}{c}{n}&\multicolumn{1}{c}{T}&\multicolumn{1}{c}{}&\multicolumn{1}{c}{1/4}&\multicolumn{1}{c}{1/2}&\multicolumn{1}{c}{3/4}&\multicolumn{1}{c}{}&\multicolumn{1}{c}{1/4}&\multicolumn{1}{c}{1/2}&\multicolumn{1}{c}{3/4}&\multicolumn{1}{c}{}&\multicolumn{1}{c}{1/4}&\multicolumn{1}{c}{1/2}&\multicolumn{1}{c}{3/4}&\multicolumn{1}{c}{}&\multicolumn{1}{c}{1/4}&\multicolumn{1}{c}{1/2}&\multicolumn{1}{c}{3/4}\tabularnewline
\hline
&&&&&&&&&&&&&&&&&\tabularnewline
$ 25$&$ 20$&&$87.4$&$87.0$&$87.7$&&$87.3$&$87.8$&$86.9$&&$84.8$&$84.7$&$74.9$&&$94.9$&$91.7$&$85.4$\tabularnewline
$ 25$&$ 50$&&$85.8$&$84.8$&$87.3$&&$85.8$&$85.4$&$87.4$&&$87.9$&$84.6$&$82.9$&&$93.4$&$90.1$&$88.7$\tabularnewline
$ 25$&$100$&&$85.1$&$85.7$&$84.9$&&$85.4$&$85.8$&$86.0$&&$85.4$&$85.5$&$80.6$&&$91.3$&$89.0$&$86.1$\tabularnewline
\hline
&&&&&&&&&&&&&&&&&\tabularnewline
$ 50$&$ 20$&&$87.7$&$88.2$&$87.3$&&$88.5$&$88.1$&$86.5$&&$86.3$&$82.5$&$77.3$&&$92.7$&$89.7$&$84.5$\tabularnewline
$ 50$&$ 50$&&$88.6$&$87.7$&$87.8$&&$88.2$&$87.2$&$88.1$&&$89.0$&$85.0$&$85.6$&&$91.9$&$89.1$&$89.5$\tabularnewline
$ 50$&$100$&&$88.5$&$87.5$&$87.7$&&$89.0$&$87.8$&$88.5$&&$87.9$&$87.5$&$84.3$&&$91.4$&$89.4$&$90.1$\tabularnewline
\hline
&&&&&&&&&&&&&&&&&\tabularnewline
$100$&$ 20$&&$89.1$&$89.9$&$88.3$&&$89.9$&$88.6$&$87.4$&&$87.4$&$84.0$&$78.1$&&$92.4$&$89.1$&$84.0$\tabularnewline
$100$&$ 50$&&$88.0$&$88.0$&$88.5$&&$87.6$&$87.9$&$87.8$&&$88.0$&$86.9$&$84.2$&&$89.7$&$89.3$&$87.2$\tabularnewline
$100$&$100$&&$86.9$&$89.6$&$89.8$&&$87.1$&$89.6$&$89.4$&&$86.1$&$88.3$&$87.7$&&$89.2$&$90.0$&$89.4$\tabularnewline
\hline
\end{tabular}\end{center}}
\end{table}

\subsubsection{Dependent case}

Now we consider the dependent case. Tables \ref{chisq_dep_loc_cover}, \ref{chisq_dep_het_cover}, and \ref{chisq_dyn_cover}, for the location, location-scale, and dynamic models, respectively.

Simulation results for coverage rates in Tables \ref{chisq_dep_loc_cover} and \ref{chisq_dep_het_cover}, considering the location and location-scale models, respectively, are similar to their corresponding previous cases. Both \textbf{RWB.p} and \textbf{RWB.se} have empirical coverage approximating well the nominal 90\%. In addition, coverages are consistent across the three quantiles, close to nominal for small samples, and improve with sample size. These approximately correct coverage results for the dependent case highlight the importance of the proposed weighted bootstrap that preserves the time series structure and allow for dependence in the data. 

The results in Tables \ref{chisq_dep_loc_cover} and \ref{chisq_dep_het_cover} also show evidence of distortions for the \textbf{RWB.t} cases. Moreover, we note that the results for the CI coverage using the asymptotic theory \textbf{AT} have larger distortions, relative to \textbf{RWB.p} and \textbf{RWB.se}, for the dependent case. Tables  \ref{chisq_dep_loc_cover} and \ref{chisq_dep_het_cover} show evidence of larger distortions for \textbf{AT}, relative to  \textbf{RWB.p} and \textbf{RWB.se}, especially for the small sample cases.

Table \ref{chisq_dyn_cover} collects the results for the dynamic case. The results show a smaller coverage for all procedures, and highlight the importance of the larger time-series in this case.  The coverage probability for the proposed methods improve substantially with sample size.

%latex.default(cover_tab, title = "Coverage prob.", file = paste0("./tables/",     file_prefix, "_cover.tex"), label = paste0(file_prefix, "_cover"),     size = "scriptsize", cgroup = c("", method_names), n.cgroup = c(2,         rep(ntaus, (ncol(cover_tab) - 2)/ntaus)), n.rgroup = rep(nT,         nn), caption = paste(desc, "Nominal 90 percent CIs.  1000 simulation\n                  repetitions used.  RWB.p, RWB.se and RWB.t are randomly\n                  weighted bootstrap confidence intervals using percentile,\n                  bootstrap standard error estimate or bootstrap t estimates\n                  respectively, and AT stands for asymptotic theory."))%
\begin{table}[!tbp]
{\scriptsize
\caption{Homoskedastic, dependent error. Nominal 90 percent CIs.  1000 simulation
                  repetitions used.  RWB.p, RWB.se and RWB.t are randomly
                  weighted bootstrap confidence intervals using percentile,
                  bootstrap standard error estimate or bootstrap t estimates
                  respectively, and AT stands for asymptotic theory.\label{chisq_dep_loc_cover}} 
\begin{center}
\begin{tabular}{rrcrrrcrrrcrrrcrrr}
\hline\hline
\multicolumn{2}{c}{\bfseries }&\multicolumn{1}{c}{\bfseries }&\multicolumn{3}{c}{\bfseries RWB.p}&\multicolumn{1}{c}{\bfseries }&\multicolumn{3}{c}{\bfseries RWB.se}&\multicolumn{1}{c}{\bfseries }&\multicolumn{3}{c}{\bfseries RWB.t}&\multicolumn{1}{c}{\bfseries }&\multicolumn{3}{c}{\bfseries AT}\tabularnewline
\cline{4-6} \cline{8-10} \cline{12-14} \cline{16-18}
\multicolumn{1}{c}{n}&\multicolumn{1}{c}{T}&\multicolumn{1}{c}{}&\multicolumn{1}{c}{1/4}&\multicolumn{1}{c}{1/2}&\multicolumn{1}{c}{3/4}&\multicolumn{1}{c}{}&\multicolumn{1}{c}{1/4}&\multicolumn{1}{c}{1/2}&\multicolumn{1}{c}{3/4}&\multicolumn{1}{c}{}&\multicolumn{1}{c}{1/4}&\multicolumn{1}{c}{1/2}&\multicolumn{1}{c}{3/4}&\multicolumn{1}{c}{}&\multicolumn{1}{c}{1/4}&\multicolumn{1}{c}{1/2}&\multicolumn{1}{c}{3/4}\tabularnewline
\hline
&&&&&&&&&&&&&&&&&\tabularnewline
$ 25$&$ 20$&&$87.0$&$85.3$&$86.6$&&$88.3$&$87.7$&$85.9$&&$83.3$&$84.7$&$75.9$&&$94.7$&$92.4$&$88.3$\tabularnewline
$ 25$&$ 50$&&$88.2$&$87.3$&$88.1$&&$88.4$&$89.5$&$88.6$&&$87.7$&$87.8$&$83.3$&&$93.8$&$93.0$&$89.2$\tabularnewline
$ 25$&$100$&&$85.9$&$84.9$&$86.7$&&$86.8$&$86.5$&$86.6$&&$86.6$&$86.0$&$84.3$&&$92.4$&$90.4$&$88.8$\tabularnewline
\hline
&&&&&&&&&&&&&&&&&\tabularnewline
$ 50$&$ 20$&&$88.6$&$87.8$&$86.7$&&$87.7$&$88.6$&$88.2$&&$86.0$&$86.3$&$81.1$&&$92.4$&$92.0$&$88.1$\tabularnewline
$ 50$&$ 50$&&$88.2$&$85.4$&$88.5$&&$87.4$&$87.1$&$88.1$&&$87.7$&$84.9$&$84.9$&&$91.7$&$89.3$&$90.1$\tabularnewline
$ 50$&$100$&&$89.2$&$86.5$&$87.9$&&$87.6$&$87.8$&$88.6$&&$86.3$&$86.9$&$86.2$&&$90.7$&$90.3$&$89.4$\tabularnewline
\hline
&&&&&&&&&&&&&&&&&\tabularnewline
$100$&$ 20$&&$86.4$&$87.8$&$89.6$&&$86.9$&$87.1$&$89.5$&&$84.4$&$83.3$&$82.7$&&$90.5$&$90.2$&$89.0$\tabularnewline
$100$&$ 50$&&$88.1$&$87.2$&$86.3$&&$87.9$&$86.5$&$86.7$&&$87.3$&$84.9$&$85.6$&&$90.5$&$88.9$&$88.5$\tabularnewline
$100$&$100$&&$89.0$&$88.9$&$90.1$&&$88.7$&$89.1$&$90.0$&&$87.0$&$88.3$&$88.1$&&$90.2$&$90.5$&$90.7$\tabularnewline
\hline
\end{tabular}\end{center}}
\end{table}

%latex.default(cover_tab, title = "Coverage prob.", file = paste0("./tables/",     file_prefix, "_cover.tex"), label = paste0(file_prefix, "_cover"),     size = "scriptsize", cgroup = c("", method_names), n.cgroup = c(2,         rep(ntaus, (ncol(cover_tab) - 2)/ntaus)), n.rgroup = rep(nT,         nn), caption = paste(desc, "Nominal 90 percent CIs.  1000 simulation\n                  repetitions used.  RWB.p, RWB.se and RWB.t are randomly\n                  weighted bootstrap confidence intervals using percentile,\n                  bootstrap standard error estimate or bootstrap t estimates\n                  respectively, and AT stands for asymptotic theory."))%
\begin{table}[!tbp]
{\scriptsize
\caption{Heteroskedastic, dependent error. Nominal 90 percent CIs.  1000 simulation
                  repetitions used.  RWB.p, RWB.se and RWB.t are randomly
                  weighted bootstrap confidence intervals using percentile,
                  bootstrap standard error estimate or bootstrap t estimates
                  respectively, and AT stands for asymptotic theory.\label{chisq_dep_het_cover}} 
\begin{center}
\begin{tabular}{rrcrrrcrrrcrrrcrrr}
\hline\hline
\multicolumn{2}{c}{\bfseries }&\multicolumn{1}{c}{\bfseries }&\multicolumn{3}{c}{\bfseries RWB.p}&\multicolumn{1}{c}{\bfseries }&\multicolumn{3}{c}{\bfseries RWB.se}&\multicolumn{1}{c}{\bfseries }&\multicolumn{3}{c}{\bfseries RWB.t}&\multicolumn{1}{c}{\bfseries }&\multicolumn{3}{c}{\bfseries AT}\tabularnewline
\cline{4-6} \cline{8-10} \cline{12-14} \cline{16-18}
\multicolumn{1}{c}{n}&\multicolumn{1}{c}{T}&\multicolumn{1}{c}{}&\multicolumn{1}{c}{1/4}&\multicolumn{1}{c}{1/2}&\multicolumn{1}{c}{3/4}&\multicolumn{1}{c}{}&\multicolumn{1}{c}{1/4}&\multicolumn{1}{c}{1/2}&\multicolumn{1}{c}{3/4}&\multicolumn{1}{c}{}&\multicolumn{1}{c}{1/4}&\multicolumn{1}{c}{1/2}&\multicolumn{1}{c}{3/4}&\multicolumn{1}{c}{}&\multicolumn{1}{c}{1/4}&\multicolumn{1}{c}{1/2}&\multicolumn{1}{c}{3/4}\tabularnewline
\hline
&&&&&&&&&&&&&&&&&\tabularnewline
$ 25$&$ 20$&&$85.6$&$86.4$&$84.3$&&$85.5$&$86.1$&$84.8$&&$82.4$&$81.9$&$73.5$&&$92.1$&$89.7$&$83.7$\tabularnewline
$ 25$&$ 50$&&$88.2$&$87.9$&$86.6$&&$88.3$&$88.0$&$85.5$&&$87.3$&$84.1$&$80.4$&&$92.5$&$90.8$&$86.7$\tabularnewline
$ 25$&$100$&&$85.5$&$86.1$&$85.6$&&$87.5$&$86.5$&$86.1$&&$86.3$&$83.3$&$83.4$&&$90.2$&$87.9$&$87.8$\tabularnewline
\hline
&&&&&&&&&&&&&&&&&\tabularnewline
$ 50$&$ 20$&&$88.4$&$88.0$&$84.5$&&$88.2$&$87.5$&$83.4$&&$82.6$&$83.4$&$74.5$&&$88.7$&$88.2$&$82.4$\tabularnewline
$ 50$&$ 50$&&$88.0$&$86.2$&$87.2$&&$86.7$&$86.3$&$86.0$&&$85.1$&$84.4$&$82.6$&&$89.1$&$87.7$&$86.0$\tabularnewline
$ 50$&$100$&&$87.9$&$87.8$&$87.8$&&$88.3$&$88.3$&$87.7$&&$86.7$&$86.1$&$84.7$&&$89.7$&$89.5$&$87.4$\tabularnewline
\hline
&&&&&&&&&&&&&&&&&\tabularnewline
$100$&$ 20$&&$86.2$&$88.3$&$81.5$&&$86.7$&$88.8$&$81.3$&&$83.7$&$83.2$&$73.3$&&$88.5$&$87.7$&$78.1$\tabularnewline
$100$&$ 50$&&$89.2$&$87.5$&$85.4$&&$88.8$&$87.3$&$84.6$&&$88.0$&$84.0$&$81.5$&&$90.2$&$87.0$&$83.6$\tabularnewline
$100$&$100$&&$89.0$&$89.8$&$89.8$&&$89.2$&$89.2$&$88.8$&&$86.2$&$86.3$&$84.9$&&$89.3$&$89.0$&$88.2$\tabularnewline
\hline
\end{tabular}\end{center}}
\end{table}

%latex.default(cover_tab, title = "Coverage prob.", file = paste0("./tables/",     file_prefix, "_cover.tex"), label = paste0(file_prefix, "_cover"),     size = "scriptsize", cgroup = c("", method_names), n.cgroup = c(2,         rep(ntaus, (ncol(cover_tab) - 2)/ntaus)), n.rgroup = rep(nT,         nn), caption = paste(desc, "Nominal 90 percent CIs.  1000 simulation\n                  repetitions used.  RWB.p, RWB.se and RWB.t are randomly\n                  weighted bootstrap confidence intervals using percentile,\n                  bootstrap standard error estimate or bootstrap t estimates\n                  respectively, and AT stands for asymptotic theory."))%
\begin{table}[!tbp]
{\scriptsize
\caption{Dynamic model. Nominal 90 percent CIs.  1000 simulation
                  repetitions used.  RWB.p, RWB.se and RWB.t are randomly
                  weighted bootstrap confidence intervals using percentile,
                  bootstrap standard error estimate or bootstrap t estimates
                  respectively, and AT stands for asymptotic theory.\label{chisq_dyn_cover}} 
\begin{center}
\begin{tabular}{rrcrrrcrrrcrrrcrrr}
\hline\hline
\multicolumn{2}{c}{\bfseries }&\multicolumn{1}{c}{\bfseries }&\multicolumn{3}{c}{\bfseries RWB.p}&\multicolumn{1}{c}{\bfseries }&\multicolumn{3}{c}{\bfseries RWB.se}&\multicolumn{1}{c}{\bfseries }&\multicolumn{3}{c}{\bfseries RWB.t}&\multicolumn{1}{c}{\bfseries }&\multicolumn{3}{c}{\bfseries AT}\tabularnewline
\cline{4-6} \cline{8-10} \cline{12-14} \cline{16-18}
\multicolumn{1}{c}{n}&\multicolumn{1}{c}{T}&\multicolumn{1}{c}{}&\multicolumn{1}{c}{1/4}&\multicolumn{1}{c}{1/2}&\multicolumn{1}{c}{3/4}&\multicolumn{1}{c}{}&\multicolumn{1}{c}{1/4}&\multicolumn{1}{c}{1/2}&\multicolumn{1}{c}{3/4}&\multicolumn{1}{c}{}&\multicolumn{1}{c}{1/4}&\multicolumn{1}{c}{1/2}&\multicolumn{1}{c}{3/4}&\multicolumn{1}{c}{}&\multicolumn{1}{c}{1/4}&\multicolumn{1}{c}{1/2}&\multicolumn{1}{c}{3/4}\tabularnewline
\hline
&&&&&&&&&&&&&&&&&\tabularnewline
$ 25$&$ 20$&&$69.4$&$61.2$&$57.8$&&$71.6$&$62.2$&$58.7$&&$79.3$&$66.6$&$51.6$&&$90.9$&$74.3$&$55.1$\tabularnewline
$ 25$&$ 50$&&$79.1$&$75.4$&$71.7$&&$79.0$&$75.6$&$73.9$&&$84.3$&$73.9$&$70.4$&&$90.7$&$79.8$&$74.0$\tabularnewline
$ 25$&$100$&&$82.1$&$81.1$&$79.6$&&$82.5$&$80.0$&$79.6$&&$84.6$&$80.3$&$78.2$&&$90.1$&$84.3$&$81.3$\tabularnewline
\hline
&&&&&&&&&&&&&&&&&\tabularnewline
$ 50$&$ 20$&&$58.9$&$43.1$&$39.4$&&$61.3$&$45.0$&$41.0$&&$70.8$&$49.0$&$34.0$&&$77.8$&$50.4$&$35.3$\tabularnewline
$ 50$&$ 50$&&$76.2$&$68.3$&$62.6$&&$75.8$&$67.7$&$64.2$&&$80.2$&$66.0$&$61.9$&&$84.7$&$70.0$&$63.8$\tabularnewline
$ 50$&$100$&&$81.6$&$77.8$&$74.1$&&$79.3$&$76.8$&$75.7$&&$80.1$&$76.1$&$73.1$&&$84.3$&$78.5$&$75.5$\tabularnewline
\hline
&&&&&&&&&&&&&&&&&\tabularnewline
$100$&$ 20$&&$37.3$&$18.2$&$15.7$&&$37.6$&$19.9$&$16.6$&&$48.0$&$22.4$&$15.2$&&$50.7$&$19.8$&$12.6$\tabularnewline
$100$&$ 50$&&$64.8$&$53.8$&$46.3$&&$64.0$&$55.3$&$45.9$&&$68.7$&$54.1$&$44.7$&&$70.8$&$55.5$&$45.4$\tabularnewline
$100$&$100$&&$77.0$&$67.9$&$64.7$&&$76.7$&$66.3$&$64.6$&&$75.6$&$63.9$&$61.7$&&$78.5$&$66.5$&$65.4$\tabularnewline
\hline
\end{tabular}\end{center}}
\end{table}

Overall the numerical results confirm that the proposed random weight bootstrap procedures are useful for practical inference in the QR panel data context. The simulation results show strong evidence that bootstrap CIs have empirical coverage close to the nominal, especially as the sample size increases.

\section{Empirical application}\label{sec:application}

\subsection{Environmental Kuznets curve}

This section illustrates the usefulness of the proposed method with an empirical example. We use panel data quantile regression methods to study the environmental Kuznets curve.
We accommodate possible heterogeneity on the effects of per capita income on the conditional distribution of environmental degradation by using FE-QR for estimation. Indeed, this heterogeneity is not revealed by conventional least squares. 

There is growing interest in environmental degradation and its relationship with economic growth, the excessive use of natural resources, and climate change. The burning of fossil fuels (i.e., carbon, oil, and gases) used for the production of energy necessary for economic development continues to significantly contribute to CO$_{2}$ emissions. The relationship between the economy and the environment is complex and controversial, and a number of empirical studies employ the Environmental Kuznets Curves (EKC) to shed light on this issue.

The EKC is based on the concept of the Kuznets curve, proposed by \citet{Kuznets55}, which describes an inverted U-shaped relationship between income inequality and income per capita. The intuition is that income inequality first increases as per capita income rises, and then starts to decrease from a certain threshold point. The notion of the Kuznets curve was applied to the environmental quality to investigate whether the relationship between income per capita and environmental degradation follows a similar inverted U-shape relationship (see, e.g., \citet{GrossmanKrueger93, GrossmanKrueger95}). In this context, the EKC postulates that low income levels are directly related to the deterioration of the environment, but after a certain level of income per capita, the this relationship reverses and becomes a negative one.\footnote{One potential rationale for the EKC would be that higher levels of development are associated with a change in economic structure in favor of industry and services which are more efficient and environmentally friendly and that, in turn, help to preserve natural resources and reduce environmental deterioration.}
  There is a relatively large more recent literature using conditional average models to estimate the EKC, see, e.g., \citet{Dinda04}; \citet{Galeotti07} and \citet{KaikaZervas13a,KaikaZervas13b} for reviews. There is no consensus about this relationship. On the one hand, \citet{SeldenSong94}, \citet{GrossmanKrueger95}, \citet{ListGallet99} and \citet{SternCommon01}, among others, find evidence in favor of inverted U-shaped relationships, at least in the case of the developed countries. On the other hand, \citet{HarbaughLevingsonWilson02} and \citet{EffiongOriabije18} argue that there is no evidence that this relationship is valid for a number of emission pollutants.

More recently, a literature using conditional quantile models investigates the EKC hypothesis. \citet{FloresFlores-LagunesKapetanakis14} employ FE-QR to estimate the relationship between economic activity and NOx (nitrogen oxide) or SO$_{2}$ (sulfur dioxide) using U.S. data.  \citet{YadumaKortelainenWossink15} apply FE-QR to estimate the EKC the CO$_{2}$ within two groups of economic development (OECD and non-OECD countries) and six geographical regions --- Western and Eastern Europe, Latin America, East Asia, West Asia and Africa.  \citet{AllardTakmanUddinAhmed18} evaluate an N-shaped EKC using panel QR, and investigate the relationship between CO$_{2}$ emissions and GDP per capita for 74 countries over the period of 1994--2012.  \citet{IkeUsmanSarkodie20} use Method of Moments QR with FE to investigate the dynamic effect of oil production on carbon emissions in 15 oil-producing countries by accounting for the role of electricity production, economic growth, democracy, and trade over the period 1980--2010.

\subsection{Model}

It is usual in the literature to use a log-linear model with a quadratic term of the affluence (per capita income) variable in line with the EKC hypothesis to capture possible existence of an inverted U-shaped relationship.\footnote{The Stochastic Impacts by Regression on Population, Affluence and Technology (STIRPAT) model for evaluating environmental change is discussed in \citet{DietzRosa97}.} The resulting panel data model is specified as follows:
\begin{equation}\label{eq:applic_model}
\ln(E_{it}) = \beta_{0} + \beta_{1} \ln(gdp_{it}) + \beta_{2} (\ln(gdp_{it}))^{2} + \beta_{3}\ln(pop_{it}) + \beta_{4} \ln(enit_{it}) + \alpha_{i} + \varepsilon_{it},
\end{equation}
where $E_{it}$ measures the environmental quality of country $i$ at time $t$; $pop$ denotes the population size; $gdp$ is the GDP per capita; and $enit$ denotes technology which is proxied by energy intensity to capture technology's damaging effect on the environment. The term $\alpha_{i}$ captures the country-specific fixed effect that is constant over time. 

The term $\varepsilon_{it}$ in equation \eqref{eq:applic_model} captures the innovation. It is usual to impose a conditional mean exogeneity to estimate the model in \eqref{eq:applic_model}. In this paper we proposed to use a QR model and impose a zero conditional quantile restriction. Hence, we estimate the following conditional quantile function:
\begin{align}
Q_{\tau}[\ln(E_{it}) | gdp_{it}, pop_{it}, enit_{it} ] &= \beta_{0}(\tau) + \beta_{1}(\tau) \ln(gdp_{it}) + \beta_{2}(\tau) (\ln(gdp_{it}))^{2} \notag \\ 
& +\beta_{3}(\tau)\ln(pop_{it}) + \beta_{4}(\tau) \ln(enit_{it}) + \alpha_{i}(\tau). \label{eq:applic_QR_model}
\end{align}
All the variables in equation \eqref{eq:applic_QR_model} are expressed in natural logarithms so the estimated coefficients are interpreted as elasticities.

The EKC conjecture could be investigated depending on the sign and statistical significance of the slope parameters of the income variable (gdp). For a given quantile $\tau$, on the one hand, if $\beta_{1}(\tau)>0$ and $\beta_{2}(\tau) =0$, then the relationship income-pollution is monotonically increasing (or decreasing if $\beta_{1}(\tau)<0$ and $\beta_{2}(\tau)=0$). On the other hand, if $\beta_{1}(\tau)>0$ and $\beta_{2}(\tau)<0$, then and inverted U-shaped curve is observed for that relationship with the turning point $E^{*}(\tau)=\frac{-\beta_{1}(\tau)}{2\beta_{2}(\tau)}$.

\subsection{Data}

Our data are taken from  \cite{SoberonDHers20}.  Summary statistics are presented in Table \ref{tbapp}.  To investigate the empirical relationship between wealth and pollution, we used panel data sets consisting of 24 OECD countries and 32 non-OECD countries for the period 1980 to 2016. Countries with insufficient data on CO$_{2}$ emissions are dropped from the database. The list of countries used for the estimation is in Appendix \ref{appB}. Using both OECD and non-OECD countries may allow to draw conclusions and compare results for both developed and developing countries.

\begin{footnotesize}
\begin{longtable} {c|cccc|cccc}
\caption{Summary Statistics}
\label{tbapp}\\
 \hline
&\multicolumn{4}{c|}{OECD}&\multicolumn{4}{c}{Non-OECD}\\
 \hline
& GDP & CO$_{2}$ & POP& ENIT & GDP & CO$_{2}$ & POP& ENIT \\
  \hline
Min. 	&	1.98	&	3.73E+05	&	814.50	&	1.72	&	0.67	&	2.48E+05	&	44.81	&	0.04	\\
1st Qu.	&	3.82	&	6.46E+06	&	11790.20	&	5.78	&	2.75	&	4.96E+06	&	871.32	&	0.87	\\
Median 	&	4.86	&	1.54E+07	&	23122.80	&	8.08	&	3.55	&	1.31E+07	&	2017.52	&	1.89	\\
Mean  	&	5.39	&	4.07E+07	&	27300.00	&	9.07	&	4.28	&	7.53E+07	&	5488.12	&	3.17	\\
3rd Qu.	&	6.72	&	5.67E+07	&	38682.80	&	10.73	&	4.74	&	4.08E+07	&	4554.75	&	4.79	\\
Max.   	&	13.34	&	3.23E+08	&	118823.60	&	27.43	&	26.29	&	1.38E+09	&	57562.53	&	18.04	\\
 \hline
\end{longtable}
\end{footnotesize}

As observed in \cite{SoberonDHers20}, the data come from two main sources. Environmental degradation captured using CO$_{2}$ emissions is obtained from the International Energy Statistics of the U.S. Energy Information Administration (EIA). CO$_{2}$ emissions (in metric tones per capita) include burning of fossil fuels and cement manufacturing, but excludes emissions from land use such as deforestation.

The data for all other variables (population, affluence, and technology) are obtained from the World Development Indicators (WDI) of the World Bank. Population (POP) is measured as total population. Affluence, which captures economic prosperity, is measured as real GDP per capita (constant 2015 US dollars). Technology is measured using energy intensity (ENIT), which is expressed as total primary energy consumption per dollar GDP (1000 BTU per year in 2015 US dollars).  %\textcolor{red}{check that the units are right}

\subsection{Results}

\begin{figure}[tbp]
\caption{Estimates and confidence intervals for EKC -- OECD countries}
\label{Fig1}\centering
\includegraphics[scale=0.51]{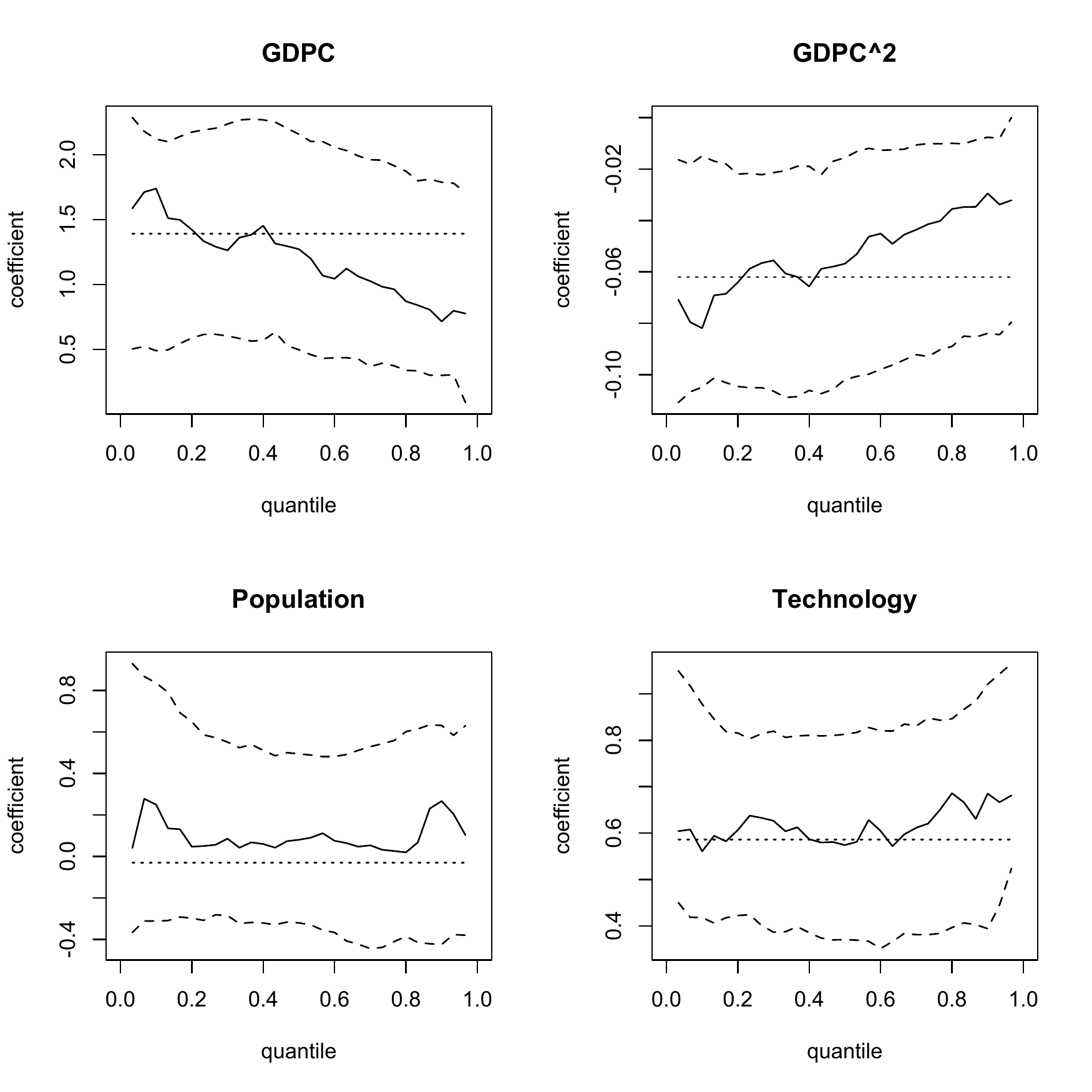}
\end{figure}

\begin{figure}[tbp]
\caption{Estimates and confidence intervals for EKC -- nonOECD countries}
\label{Fig2}\centering
\includegraphics[scale=0.51]{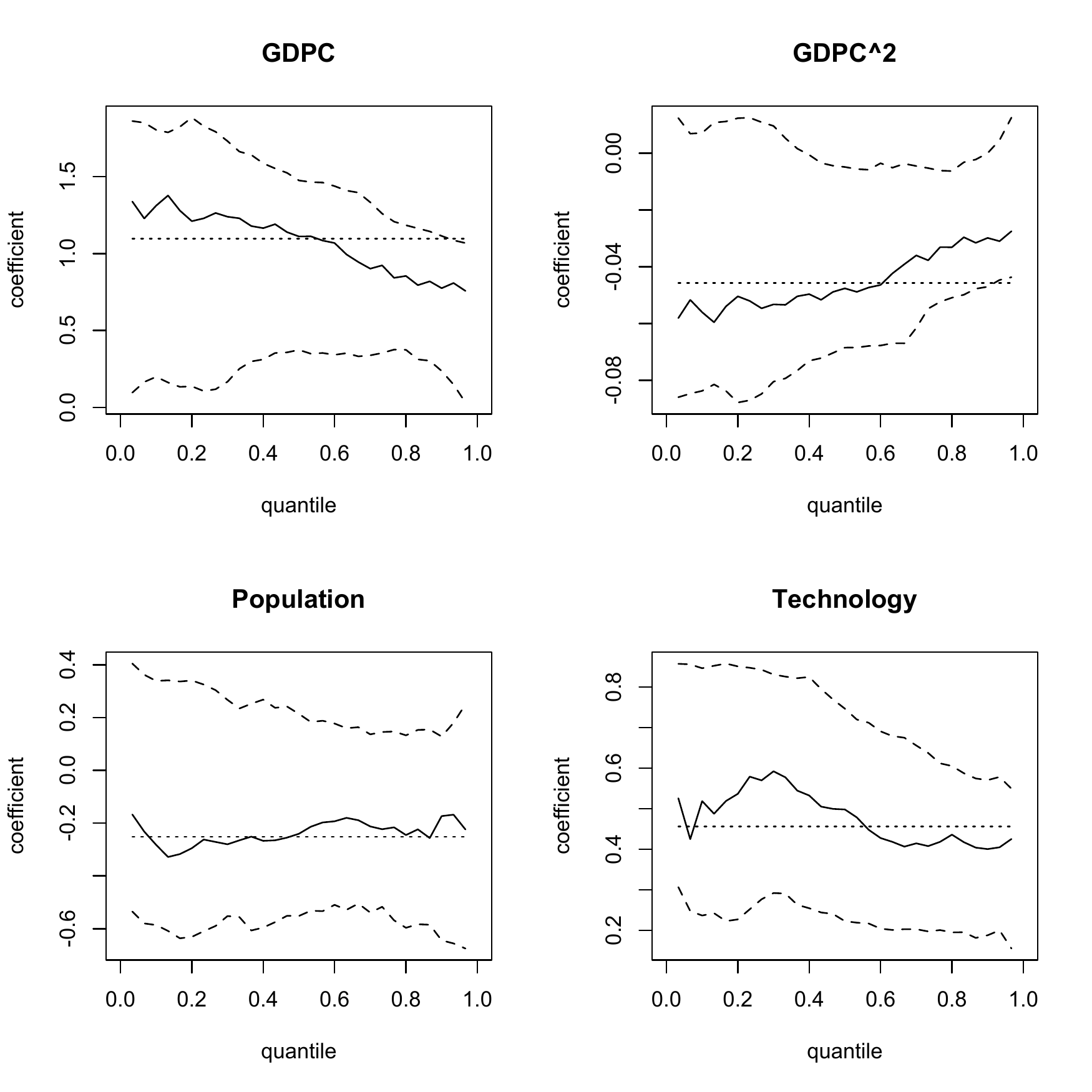}
\end{figure}

The results for the coefficient estimates across the quantiles are reported in Figures \ref{Fig1} and \ref{Fig2} for OECD and non-OECD, respectively.

First, consider the GDP coefficient in Figure \ref{Fig1}. This is strong evidence that the coefficient is positive and statistically different from zero across all conditional quantiles. In addition, it is is decreasing across quantiles. Thus, the income-pollution relationship is monotonically decreasing across conditional quantiles of environmental quality. Recall that the dependent variable measures environmental degradation, measured by CO$_{2}$ emissions. Thus, for low quantiles of the conditional distribution environmental degradation, an increase on GDP has a relatively larger impact. 

Regarding the EKC hypothesis --- a positive coefficient on GDP and negative coefficient on GDP$^{2}$ --- Figure \ref{Fig1} displays evidence that supports the EKC hypothesis for the OECD countries. Note that the coefficient estimates $\widehat{\beta}_{1}(\tau)$ are statistically positive across quantiles and $\widehat{\beta}_{2}(\tau)$ are statistically negative across quantiles as well, hence the evidence of  inverted U-shaped curve is observed across the quantiles. Interestingly, $\widehat{\beta}_{1}(\tau)$ is decreasing and $\widehat{\beta}_{2}(\tau)$ increasing across quantiles, such that the ratio $-\widehat{\beta}_{1}(\tau)/2\widehat{\beta}_{2}(\tau)$ is slightly increasing. That is, the shape of the curve is flatter for those countries that have higher emissions conditional on population, energy intensity and country-specific fixed effects.  Hence, the turning point $E^{*}$ is larger at the top of the conditional distribution of environmental degradation.

The empirical estimates for the energy intensity variable are positive, statistically significant, and slightly decreasing across quantiles for the OECD countries in Figure \ref{Fig1}. This implies that, for a given quantile, higher consumption of fossil fuels in the production process is associated with increased CO$_{2}$ emissions that in turn increase pressure on environmental quality. But this coefficient is slightly larger for larger quantiles of the conditional distribution of CO$_{2}$. For the population variable, the point estimates are negative across quantiles, but the confidence intervals are quite large and the estimates are statistically indistinguishable from zero.

Figure \ref{Fig2} displays the results for the non-OECD countries. The income-pollution is monotonically decreasing across quantiles. However, one cannot reject the hypothesis that the coefficients for GDP$^{2}$ are statistically equal to zero. Hence, the EKC hypothesis is not empirically valid for non-OECD countries, and we can only observe a positive relationship between income and pollution. The estimates regarding the energy intensity variable for non-OECD countries are also positive and statistically different from zero. They are small when compared to the OECD countries in Figure \ref{Fig1}.

The FE-QR method allow us to estimate the impacts of GDPC, population, and technology on the EKC at different quantiles of the conditional distribution of environmental degradation, as well as investigate the empirical validity of the EKC hypothesis. Our empirical findings document the validity of the income-pollution relationship, which is diminishing along the conditional quantiles of degradation for both OECD and non-OECD countries. There is empirical evidence supporting the EKC hypothesis for OECD countries, but not for non-OECD.

\section{Conclusion}\label{sec:conclusion}

This paper develops bootstrap inference methods for panel data quantile regression models with fixed effects. We consider the case of randomly-weighted bootstrap  and propose to construct asymptotically valid bootstrap standard errors, confidence intervals, and hypothesis tests for the parameters of interest using this resampling technique. The weighted bootstrap method has the advantage that it does not require estimating the variance-covariance matrix, which allows one to circumvent the need to select a bandwidth for conditional density estimation. The weights are only a function of the cross-section dimension such that the serial correlation in the original data can be preserved, thus encompassing a large class of possible empirical applications. We formally establish the asymptotic validity of the randomly-weighted bootstrap by showing consistency of the bootstrap method in distribution and consistency of bootstrap covariance estimation.

The bootstrap algorithm is simple to implement in practice. Monte Carlo simulations confirm that the proposed methods have correct finite sample properties. Numerical simulations also show evidence that confidence intervals based on the asymptotic normal approximation can be very distorted in finite samples. Instead, the proposed bootstrap greatly reduces these distortions. Finally, we provide an empirical illustration using the environmental Kuznets curve.

\newpage \appendix

%\singlespacing

\section{Proofs}

On notation: For a process $W(f)$ defined on $f\in \mathcal{F}$, where $\mathcal{F}$ is a class of measurable functions on measure space $($S$, \mathcal{S})$, let $\left\Vert W(f)\right\Vert _{\mathcal{F}}:=\sup_{f\in \mathcal{F}}\left\vert W(f)\right\vert$. For a probability measure $Q$ on $($S$,\mathcal{S})$, and $\epsilon >0$, let $N\left( \mathcal{F},L_{p}(Q),\epsilon \right) $ denote the $\epsilon $-covering number of $\mathcal{F}$ w.r.t the $L_{p}(Q)$ norm $\left\Vert \cdot \right\Vert _{L_{p}(Q)}$.  For notational simplicity, we suppress dependence on $ \tau $, and denote, for example, $(\widehat{\alpha}_{i\tau}^{\ast}, \widehat{\bm{\beta}}_\tau^{\ast})$ by $(\widehat{\alpha}_{i}^{\ast}, \widehat{\bm{\beta}}^{\ast})$.  For probabilities conditional on the observations we use starred versions of the usual notation, so for example, $\E^{\ast} [Z^{\ast}]$ means the expected value of $Z^{\ast}$ conditional on the data, and $Z_n^{\ast} \pconvs Z$ and $Z_n^{\ast} \dconvs Z$ mean that the sequence $\{Z_n^{\ast} \}$ converges to $Z$ in probability or in distribution respectively conditional on the data.  Let $\bm{\alpha} = (\alpha_1, \ldots \alpha_n)\tr$ and $\bm{\theta} =(\bm{\alpha}\tr,\bm{\beta}\tr)\tr$ and for $i = 1, \ldots, n$, let $\bm{\theta}_i = (\alpha_i, \bm{\beta}\tr)\tr$.  Also define the extended covariate vector $\mathbf{z}_{it} = (1, \mathbf{x}_{it}\tr)\tr$ so that, for example, we may write $y_{it} - \alpha_i - \mathbf{x}_{it}\tr \bm{\beta} = y_{it} - \mathbf{z}_{it}\tr \bm{\theta}_i$.  Let the conditional PDF and CDF of $y_{i1} - Q_{y_{i1} | \mathbf{z}_{i1}}(\tau | \mathbf{z}_{i1})$ be $f_i(y | \mathbf{z}_{i1}) := f_{y_{i1} | \mathbf{z}_{i1}}(y + Q_{y_{i1}|\mathbf{z}_{it}}(\tau | \mathbf{z}_{it}) | \mathbf{z}_{i1})$ and $F_i(y | \mathbf{z}_{i1}) := F_{y_{i1} | \mathbf{z}_{i1}}(y + Q_{y_{i1}|\mathbf{z}_{i1}}(\tau | \mathbf{z}_{i1}) | \mathbf{z}_{i1})$ respectively (note that $f_i(y)$, the corresponding unconditional density, was defined in the main text).  The sample and resampled objective functions are defined as
\begin{align}
  L_{n}(\bm{\theta}) &:= \frac{1}{nT}\sum_{i=1}^{n} \sum_{t=1}^{T} \rho_{\tau} (y_{it} - \mathbf{z}_{it}\tr\bm{\theta}_i) := \frac{1}{n} \sum_{i=1}^{n} L_{ni}(\bm{\theta}_i) \notag \\
  \intertext{and}
  L_{n}^{\ast} (\bm{\theta}) &:= \frac{1}{nT} \sum_{i=1}^{n} \omega_{i}\sum_{t=1}^{T}\rho _{\tau }(y_{it} - \mathbf{z}_{it}\tr \bm{\theta}_i) := \frac{1}{n} \sum_{i=1}^{n} L_{ni}^{\ast} (\bm{\theta}_i). \label{lstar} 
\end{align}

We first provide three auxiliary results to demonstrate Theorem \ref{thm:consistent}.

\begin{lemma}[Consistency] \label{lem:consistent_est}
  Assume that $n / T \to 0$ as $n \to \infty$ and that $T$ grows at most polynomially in $n$.  Then under conditions \ref{a:data}-\ref{a:minf}, $\max_{1 \leq i \leq n} | \widehat{\alpha}_i^{\ast} - \widehat{\alpha}_i | \pconvs 0$ and $\widehat{\bm{\beta}}^{\ast} \pconvs \widehat{\bm{\beta}}$.
\end{lemma}

\begin{proof}[Proof of Lemma~\ref{lem:consistent_est}]
  Start by considering consistency of $\widehat{\bm{\beta}}^{\ast}$.  Fix a $\delta > 0$ and define the $\delta$-ball around $\widehat{\bm{\theta}}_i$, $B_i(\delta) = \{ \bm{\theta}_i \in \R^{p+1} : \|\bm{\theta}_i - \widehat{\bm{\theta}}_i\|_1 \leq \delta\}$. Denote its boundary by $\partial B_i(\delta)$.  For any $\bm{\theta}_i \not\in B_i(\delta)$, let $r_i = \delta \|\bm{\theta} - \widehat{\bm{\theta}}\|_1^{-1}$ and the convex combination $\bar{\bm{\theta}}_i = r_i \bm{\theta}_i + (1 - r_i) \widehat{\bm{\theta}}_i$.  Let $\calL^{\ast}_{ni}(\bm{\theta}_i) = L^{\ast}_{ni}(\bm{\theta}_i) - L^{\ast}_{ni}(\widehat{\bm{\theta}}_i)$ and note that $\frac{1}{n} \sum_i \calL^{\ast}_{ni}(\bm{\theta})$ is minimized at $\widehat{\bm{\theta}}_i$.  By the convexity of $L^{\ast}_{ni}$, for any $\bm{\theta}_i \notin B_i(\delta)$ we have
  \begin{align*}
    r_i( L^{\ast}_{ni}(\bm{\theta}_i) - L^{\ast}_{ni}(\widehat{\bm{\theta}}_i) ) &\geq L^{\ast}_{ni}(\bar{\bm{\theta}}) - L^{\ast}_{ni}(\widehat{\bm{\theta}}_i) \\
    {} &= \E^{\ast}[\calL^{\ast}_{ni}(\bar{\bm{\theta}}_i)] + (\calL^{\ast}_{ni}(\bar{\bm{\theta}}_i) - \E^{\ast}[\calL^{\ast}_{ni}(\bar{\bm{\theta}}_i)]).
  \end{align*}

  Now we show that given $\delta$, there is an $\epsilon_\delta > 0$ such that $\E^{\ast}[\calL^{\ast}_{ni}(\bar{\bm{\theta}}_i)] \geq \epsilon_\delta$ for all $i$.  Knight's identity \citep{Knight98} implies, with $\widehat{e}_{it} = y_{it} - \mathbf{z}_{it}^\top \widehat{\bm{\theta}}_i$,
  \begin{align*}
    \E^{\ast} \left[ \calL^{\ast}_{ni}(\bm{\theta}_i) \right] &= \frac{1}{T} \sum_{t=1}^T \left( \rho_\tau \left( \widehat{e}_{it} - \mathbf{z}_{it}\tr (\bm{\theta}_i - \widehat{\bm{\theta}}_i) \right) - \rho_\tau(\widehat{e}_{it}) \right) \\
    {} &= -\frac{1}{T} \sum_{t=1}^T \mathbf{z}_{it} \psi_\tau(\widehat{e}_{it}) + \frac{1}{T} \sum_{t=1}^T \int_0^{ \mathbf{z}_{it}\tr (\bm{\theta}_i - \widehat{\bm{\theta}}_i) } I(\widehat{e}_{it} \leq s) - I(\widehat{e}_{it} \leq 0) \ud s.
  \end{align*}
  Lemma~\ref{lem:subg} below implies that $\| \frac{1}{T} \sum_{t=1}^T \mathbf{z}_{it} \psi_\tau(\widehat{e}_{it}) \| = O(n / T)$ almost surely (assuming that $n + p < T$).  Then we can rewrite, with $e_{it} = y_{it} - \mathbf{z}_{it}^\top \bm{\theta}_{i0}$,
  \begin{align*}
    \E^{\ast} \left[ \calL^{\ast}_{ni}(\bm{\theta}_i) \right] &= \frac{1}{T} \sum_{t=1}^T \int_0^{ \mathbf{z}_{it}\tr (\bm{\theta}_i - \widehat{\bm{\theta}}_i) } \big\{ I(e_{it} \leq s + \mathbf{z}_{it}\tr (\widehat{\bm{\theta}}_i - \bm{\theta}_{i0})) \\
    {} &\phantom{=} \qquad \qquad - I(e_{it} \leq \mathbf{z}_{it}\tr (\widehat{\bm{\theta}}_i - \bm{\theta}_{i0})) \big\} \ud s + O(n / T) \\
    {} &= \frac{1}{T} \sum_{t=1}^T \int_0^{ \mathbf{z}_{it}\tr (\bm{\theta}_i - \widehat{\bm{\theta}}_i) } \big\{ F_i (s + \mathbf{z}_{it}\tr (\widehat{\bm{\theta}}_i - \bm{\theta}_{i0}) | \mathbf{z}_{it}) \\
    {} &\phantom{=} \qquad \qquad - F_i ( \mathbf{z}_{it}\tr (\widehat{\bm{\theta}}_i - \bm{\theta}_{i0}) | \mathbf{z}_{it} ) \big\} \ud s + O(n / T) + O_p( T^{-3/4} (\log n)^{5/4} ) \\
    {} &= \frac{1}{T} \sum_{t=1}^T f_i ( \mathbf{z}_{it}\tr (\widehat{\bm{\theta}}_i - \bm{\theta}_{i0}) | \mathbf{z}_{it} ) \int_0^{\mathbf{z}_{it}\tr (\bm{\theta}_i - \widehat{\bm{\theta}}_i)} s \ud s \\
    {} &\phantom{=} \qquad \qquad \qquad + O_p( \| \widehat{\bm{\theta}}_i - \bm{\theta}_{i0} \|^2) + O(n / T) + O_p( T^{-3/4} (\log n)^{5/4} ) \\
    {} &= ( \bm{\theta}_i - \widehat{\bm{\theta}}_i ) \frac{1}{2T} \sum_{t=1}^T f_i ( \mathbf{z}_{it}\tr (\widehat{\bm{\theta}}_i - \bm{\theta}_{i0}) | \mathbf{z}_{it} ) \mathbf{z}_{it} \mathbf{z}_{it}\tr ( \bm{\theta}_i - \widehat{\bm{\theta}}_i ) \\
    {} &\phantom{=} \qquad \qquad \qquad + O(n / T) + O_p( T^{-3/4} (\log n)^{5/4} ).
  \end{align*}
  The second equality above is due to Lemma 5 of \citet{GalvaoGuVolgushev20}, and uses Theorem~5.1 of \citet{KatoGalvaoMontes-Rojas12}, which shows that $\|\widehat{\bm{\theta}}_i - \bm{\theta}_{i0}\| = O_p(T^{-1/2} (\log n)^{1/2})$ for all $i$, along with the assumption that $T$ grows at most polynomially in $n$.  Then under Assumption~\ref{a:minf} and for large $n,T$, we can find $\epsilon_\delta > 0$ such that $\E^{\ast}[\calL^{\ast}_{ni}(\bar{\bm{\theta}}_i)] \geq \epsilon_\delta$ for all $i$ since the main term is a positive definite quadratic form.

  As in Theorem 3.1 of \citet{KatoGalvaoMontes-Rojas12}, given the above
  \begin{equation*}
    \left\{ \| \widehat{\bm{\beta}}^{\ast} - \widehat{\bm{\beta}} \|_1 > \delta \right\} \subseteq \left\{ \max_{i \in \{1, \ldots n\}} \sup_{\bm{\theta}_i \in B_i(\delta)} | \calL^{\ast}_{ni}(\bm{\theta}_i) - \E^{\ast}[\calL^{\ast}_{ni}(\bm{\theta}_i)] | > \epsilon_\delta \right\},
  \end{equation*}
  and following their argument further (taking note of their Remark 3.1 and our condition~\ref{a:boundedZ}) implies that for all $\epsilon > 0$,
  \begin{equation} \label{beta_steq}
    \lim_{n \rightarrow \infty} \P^{\ast} \left\{ \max_{i \in \{1, \ldots n\}} \sup_{\bm{\theta}_i \in B_i(\delta)} | \calL^{\ast}_{ni}(\bm{\theta}_i) - \E^{\ast}[\calL^{\ast}_{ni}(\bm{\theta}_i)] | > \epsilon \right\} = 0,
  \end{equation}
  which implies $\widehat{\bm{\beta}}^{\ast} \pconvs \widehat{\bm{\beta}}$.

  Next we turn to $\widehat{\bm{\alpha}}^{\ast}$.  Similarly to the argument for $\widehat{\bm{\beta}}^{\ast}$, recall that for each $i$, $\widehat{\alpha}_i^{\ast} = \argmin_{\alpha_i} L_{ni}^{\ast}(\alpha_i, \widehat{\bm{\beta}}^{\ast})$.  Fix a $\delta > 0$.  For any $\alpha_i$ such that $|\alpha_i - \widehat{\alpha}_i| > \delta$, let $\widetilde{r}_i = \delta / |\alpha_i - \widehat{\alpha}_i|$ and $\widetilde{\alpha}_i = \widetilde{r}_i \alpha_i + (1 - \widetilde{r}_i) \widehat{\alpha}_i$.  Convexity implies that
  \begin{equation*}
    \widetilde{r}_iL_{ni}^{\ast}(\alpha_i, \widehat{\bm{\beta}}^{\ast}) - (1 - \widetilde{r}_i) L_{ni}^{\ast}(\widehat{\alpha}_i, \widehat{\bm{\beta}}^{\ast}) \geq L_{ni}^{\ast}(\widetilde{\alpha}_i, \widehat{\bm{\beta}}^{\ast}) - L_{ni}^{\ast}(\widehat{\alpha}_i, \widehat{\bm{\beta}}^{\ast}).
  \end{equation*}
  The right-hand side can be rewritten (noting that $\E^{\ast}[\calL_{ni}^{\ast}(\widehat{\alpha}_i, \widehat{\bm{\beta}})] = 0$) as
  \begin{align*}
    L_{ni}^{\ast}(\widetilde{\alpha}_i, \widehat{\bm{\beta}}^{\ast}) - L_{ni}^{\ast}(\widehat{\alpha}_i, \widehat{\bm{\beta}}) &- (L_{ni}^{\ast}(\widehat{\alpha}_i, \widehat{\bm{\beta}}^{\ast}) - L_{ni}^{\ast}(\widehat{\alpha}_i, \widehat{\bm{\beta}}) \\
    {} &= \calL_{ni}^{\ast}(\widetilde{\alpha}_i, \widehat{\bm{\beta}}^{\ast}) - \calL_{ni}^{\ast}(\widehat{\alpha}_i, \widehat{\bm{\beta}}^{\ast}) \\
    {} &= (\calL_{ni}^{\ast}(\widetilde{\alpha}_i, \widehat{\bm{\beta}}^{\ast}) - \E^{\ast}[\calL_{ni}^{\ast}(\widetilde{\alpha}_i, \bm{\beta})]|_{\bm{\beta} = \widehat{\bm{\beta}}^{\ast}}) \\
    {} &- (\calL_{ni}^{\ast}(\widehat{\alpha}_i, \widehat{\bm{\beta}}^{\ast}) - \E^{\ast}[\calL_{ni}^{\ast}(\widehat{\alpha}_i, \bm{\beta})]|_{\bm{\beta} = \widehat{\bm{\beta}}^{\ast}}) \\
    {} &+ \E^{\ast}[\calL_{ni}^{\ast}(\widetilde{\alpha}_i, \widehat{\bm{\beta}})] \\
    {} &+ (\E^{\ast}[\calL_{ni}^{\ast}(\widetilde{\alpha}_i, \bm{\beta})]|_{\bm{\beta} = \widehat{\bm{\beta}}^{\ast}}) - \E^{\ast}[\calL_{ni}^{\ast}(\widetilde{\alpha}_i, \widehat{\bm{\beta}})]) \\
    {} &+ (\E^{\ast}[\calL_{ni}^{\ast}(\widehat{\alpha}_i, \bm{\beta})]|_{\bm{\beta} = \widehat{\bm{\beta}}^{\ast}}) - \E^{\ast}[\calL_{ni}^{\ast}(\widehat{\alpha}_i, \widehat{\bm{\beta}})]).
  \end{align*}
  Once again, $\E^{\ast}[\calL_{ni}^{\ast}(\widetilde{\alpha}_i, \widehat{\bm{\beta}})] \geq \epsilon_\delta$ for all $i$ for large enough $T$.  Then, we have the inclusion $\left\{ \exists i \text{ s.t. } |\widehat{\alpha}_i^{\ast} - \widehat{\alpha}_i| > \delta \right\} \subset A_{1n}^{\ast} \cup A_{2n}^{\ast}$, where
  \begin{align*}
    A_{1n}^{\ast} &= \left\{ \sup_i \sup_{|\alpha_i - \widehat{\alpha}_i| \leq \delta} \left| \calL_{ni}^{\ast}(\alpha_i, \widehat{\bm{\beta}}^{\ast}) - \E^{\ast}[ \calL_{ni}^{\ast}(\alpha_i, \bm{\beta})]|_{\bm{\beta} = \widehat{\bm{\beta}}^{\ast}} \right| > \epsilon_\delta / 4 \right\} \\
    \intertext{and}
    A_{2n}^{\ast} &= \left\{ \sup_i \sup_{|\alpha_i - \widehat{\alpha}_i| \leq \delta} \left| \E^{\ast}[ \calL_{ni}^{\ast}(\alpha_i, \bm{\beta})]|_{\bm{\beta} = \widehat{\bm{\beta}}^{\ast}} - \E^{\ast}[ \calL_{ni}^{\ast}(\alpha_i, \widehat{\bm{\beta}}) ] \right| > \epsilon_\delta / 4 \right\}.
  \end{align*}
  Then $\widehat{\bm{\beta}}^{\ast} \pconvs \widehat{\bm{\beta}}$ and~\eqref{beta_steq} imply that $\P^{\ast} \{ A_{1n}^{\ast} \} \pconvs 0$.  Next, because $|\E^{\ast} [ \calL_{ni}^{\ast}(\alpha_i, \bm{\beta}) ]|_{\bm{\beta} = \widehat{\bm{\beta}}^{\ast}} - \E^{\ast} [\calL_{ni}^{\ast}(\alpha_i, \widehat{\bm{\beta}})] | \leq 2(1 + M) \|\widehat{\bm{\beta}}^{\ast} - \widehat{\bm{\beta}} \| = o_{p^{\ast}}(1)$, we also have $\P^{\ast}\{ A_{2n}^{\ast} \} \pconvs 0$.  This implies that $\sup_i | \widehat{\alpha}_i^{\ast} - \widehat{\alpha}_i | \pconvs 0$.
\end{proof}

The following lemma assumes that the data are in general position, that is, under the assumption that the regression's solution as a linear program passes through exactly $n + p$ design points \citep[Definition 2.1]{Koenker05}.  This is the case almost surely under Assumptions~\ref{a:boundedf} and~\ref{a:minf}.  Note that the unperturbed solution to the problem may be characterized by setting $\omega_i = 1$ for all $i$.

\begin{lemma}[Subgradient condition] \label{lem:subg}
  Suppose that $\widehat{\bm{\theta}}^\ast$ is the minimizer of the objective function $L_n^\ast(\bm{\theta})$ defined in~\eqref{lstar}.  Suppose that the data are in general position, that is, assume that any $n + p$ observations result in a unique, exact fit.  Then with $\psi_\tau(u) = \tau - I(u \leq 0)$,
  \begin{equation*}
    \left| -\frac{\omega_i}{T} \sum_{t=1}^{T} \psi_\tau (y_{it} - \mathbf{z}_{it}^{\top} \widehat{\bm{\theta}}_i^\ast ) \right| \leq \frac{\min \{(n + p), T\}}{T} \omega_i, \quad i = 1, \ldots n.
  \end{equation*}
  Assuming that $n + p < nT$,
  \begin{equation*}
    \left| -\frac{1}{nT} \sum_{i=1}^n \omega_i \sum_{t=1}^T \psi_\tau (y_{it} - \mathbf{z}_{it}^{\top} \widehat{\bm{\theta}}_i^\ast ) \right| \leq \frac{n + p}{nT} \sup_i \omega_i.
  \end{equation*}
  Also assuming that $n + p < nT$,
  \begin{equation*}
    \left\| -\frac{1}{nT} \sum_{i=1}^{n} \omega_i \sum_{t=1}^{T} \mathbf{x}_{it} \psi_{\tau} ( y_{it} - \mathbf{z}_{it}^{\top} \widehat{\bm{\theta}}_i^\ast ) \right\| \leq \frac{n + p}{nT} \sup_i \omega_i \max_{i,t} \|\mathbf{x}_{it}\|.
  \end{equation*}
\end{lemma}

\begin{proof}[Proof of Lemma~\ref{lem:subg}]
  Define the unperturbed design matrix $\bm{W} \in \mathbb{R}^{nT \times (p + T)}$ by $\bm{W} = [\begin{smallmatrix} \bm{X} & I_n \otimes \mathbf{1}_T \end{smallmatrix}]$, where $\bm{X}$ is the matrix with rows $\bm{x}_{it}^\top$ for all $i$ and $t$.  Then let $\bm{\Omega}$ be an $nT \times nT$ diagonal matrix with $T$ contiguous copies of each realized $\omega_i$ along the diagonal.  Then let $\bm{W}^\ast = \bm{\Omega} \bm{W}$ and $\bm{y}^\ast = \bm{\Omega} \bm{y}$ (where $\bm{y}$ is the vector of all $y_{it}$).  As the solution to a linear programming problem in general position, the QR solution is a linear combination of $n + p$ basic $(y_{it}^\ast, \bm{x}_{it}^\ast)$ observations.  Let $\bm{h}$ index these basic observations out of the $nT$ total observations and write $\bm{W}^\ast(\bm{h})$ to denote the $(n + p) \times (n + p)$ submatrix of $\bm{W}^\ast$ with rows corresponding to the basic observations (so then $\widehat{\bm{\theta}} = \bm{W}^\ast(\bm{h})^{-1} \bm{y}^\ast(\bm{h})$).  \citet[Equation 3.10]{GutenbrunnerJureckova92} or the calculations leading to Theorem 2.1 of \citet{Koenker05} imply that, with the inequalities interpreted in a coordinate-wise manner,
    \begin{equation} \label{dual_ineq}
      -\tau \mathbf{1}_{n + p}^\top \bm{W}^\ast(\bm{h}) \leq \sum_{i=1}^n \sum_{t=1}^T \bm{w}_{it}^{\ast\top} \psi_\tau(\widehat{e}_{it}^\ast) \leq (1 - \tau) \mathbf{1}_{n + p}^\top \bm{W}^\ast(\bm{h}),
  \end{equation}
  where $\widehat{e}_{it}^\ast = y_{it} - \widehat{\alpha}_i^\ast - \bm{x}_{it}^{\ast\top} \widehat{\bm{\beta}}^\ast$.

  Define a selection vector $\xi_i = [\begin{smallmatrix} \mathbf{0}_p \\ \bm{e}_i \end{smallmatrix}]$, where $\bm{e}_i$ is the $i$th standard basis vector in $\mathbb{R}^n$.  Note that $-\frac{\omega_i}{T} \sum_{t=1}^{T} \psi_\tau (\widehat{e}_{it}^\ast ) = -\frac{1}{T} \sum_{i,t} \bm{w}_{it}^{\ast\top} \psi_\tau(\widehat{e}_{it}^\ast) \xi_i$.  Postmultiply all parts of~\eqref{dual_ineq} by $\xi_i$ and note that at most $\min\{(n + p), T\}$ basic observations come from unit $i$ to find the first result.

    Next, postmultiply the terms in~\eqref{dual_ineq} by $[\begin{smallmatrix} \mathbf{0}_p \\ \mathbf{1}_n \end{smallmatrix}]$, recalling that there are $n + p < nT$ basic observations total.  This implies that, letting $\{\tilde{\omega}_k\}_{k=1}^{n + p}$ be some selection, with replacement, from $\{\omega_i\}_{i=1}^n$,
  \begin{equation*}
    \left| -\frac{1}{nT} \sum_{i=1}^n \omega_i \sum_{t=1}^T \psi_\tau (\widehat{e}_{it}^\ast) \right| \leq \frac{1}{nT} \sum_{k=1}^{n + p} \tilde{\omega}_k \leq \frac{n + p}{nT} \sup_i \omega_i.
  \end{equation*}

    Finally define a selection matrix $\Xi = [\begin{smallmatrix} I_p \\ \mathbf{0}_{T \times p} \end{smallmatrix}]$.  Similarly, that $\|-\frac{1}{nT} \sum_{i=1}^{n} \omega_i \sum_{t=1}^{T} \mathbf{x}_{it} \psi_{\tau} ( \widehat{e}_{it}^\ast ) \| = \| \frac{1}{nT} \sum_{i,t} \bm{w}_{it}^{\ast\top} \psi_\tau(\widehat{e}_{it}^\ast) \Xi \|$ so postmultiplication by $\Xi$ implies the final result using calculations similar to those in the previous part.
\end{proof}

\begin{lemma}[Asymptotic normality] \label{lem:Anormal}
  Assume that $T= O(n^{r})$ for some $r > 0$ and $n (\log T)^{4}/T \to 0$ as $n \to \infty$. Then, under conditions \ref{a:data}-\ref{a:minf}, \ref{a:ydens}, \ref{a:bweight}, and \ref{a:Avar}
\begin{equation*}
  \sqrt{nT} ( \widehat{\bm{\beta}}^{\ast} - \widehat{\bm{\beta}} ) \dconvs \mathcal{N} (\bm{0}, \Gamma^{-1} V \Gamma^{-1}).
\end{equation*}
\end{lemma}

\begin{proof}[Proof of Lemma~\ref{lem:Anormal}]
  Define
\begin{align*}
  \mathbb{H}_{Ti} (\bm{\theta}_i) &:= -\frac{1}{T} \sum_{t=1}^{T} \psi_{\tau} \left( e_{it} - \mathbf{z}_{it}^\top \left( \bm{\theta}_i - \bm{\theta}_{i0} \right) \right), \quad i = 1, \ldots, n, \\
  H_{Ti} (\bm{\theta}_i) &:= -\frac{1}{T} \sum_{t=1}^{T} \left( \tau - F_i \left( \mathbf{z}_{it}^\top \left( \bm{\theta}_i - \bm{\theta}_{i0} \right) | \mathbf{z}_{it} \right) \right), \quad i = 1, \ldots, n, \\
  \mathbb{S}_{nT}^{\ast} (\bm{\theta}) &:= -\frac{1}{nT} \sum_{i=1}^n \omega_i \sum_{t=1}^T \mathbf{x}_{it} \psi_{\tau} \left( e_{it} - \mathbf{z}_{it}^\top \left( \bm{\theta}_i - \bm{\theta}_{i0} \right) \right), \\
  S_{nT}^{\ast} (\bm{\theta}) &:= -\frac{1}{nT} \sum_{i=1}^n \omega_i \sum_{t=1}^T \mathbf{x}_{it} \left( \tau - F_i \left( \mathbf{z}_{it}^\top \left( \bm{\theta}_i - \bm{\theta}_{i0} \right) | \mathbf{z}_{it} \right) \right).
\end{align*}

For each $i$ write
\begin{equation} \label{alpha_expansion}
  \mathbb{H}_{Ti} (\widehat{\bm{\theta}}_i^{\ast}) = \mathbb{H}_{Ti} (\widehat{\bm{\theta}}_i) + \left( H_{Ti} (\widehat{\bm{\theta}}_i^{\ast}) - H_{Ti} (\widehat{\bm{\theta}}_i) \right) + \mathcal{H}_{Ti} (\widehat{\bm{\theta}}_i^{\ast}, \widehat{\bm{\theta}}_i)
\end{equation}
where
\begin{equation*}
  \mathcal{H}_{Ti}(\bm{\theta}_i, \bm{\theta}_i') := \mathbb{H}_{Ti} (\bm{\theta}_i) - \mathbb{H}_{Ti} (\bm{\theta}_i) - H_{Ti} (\bm{\theta}_i') + H_{Ti} (\bm{\theta}_i').
\end{equation*}
Rewrite
\begin{align}
  H_{Ti} (\widehat{\bm{\theta}}_i^{\ast}) - H_{Ti} (\widehat{\bm{\theta}}_i) &= \frac{1}{T} \sum_{t=1}^T \left( F_i ( \mathbf{z}_{it}\tr (\widehat{\bm{\theta}}_i^{\ast} - \bm{\theta}_{i0}) | \mathbf{z}_{it} ) - F_i ( \mathbf{z}_{it}\tr (\widehat{\bm{\theta}}_i - \bm{\theta}_{i0}) | \mathbf{z}_{it} ) \right) \notag \\
  {} &= \frac{1}{T} \sum_{t=1}^T f_i ( \mathbf{z}_{it}\tr (\widehat{\bm{\theta}}_i - \bm{\theta}_{i0}) | \mathbf{z}_{it} ) (\widehat{\alpha}_i^{\ast} - \widehat{\alpha}_{i}) \notag \\
  {} &\phantom{=} \quad + \frac{1}{T} \sum_{t=1}^T f_i ( \mathbf{z}_{it}\tr (\widehat{\bm{\theta}}_i - \bm{\theta}_{i0}) | \mathbf{z}_{it} ) \mathbf{x}_{it}\tr (\widehat{\bm{\beta}}^{\ast} - \widehat{\bm{\beta}}) + O_{p^{\ast}}( \| \widehat{\bm{\theta}}_i^{\ast} - \widehat{\bm{\theta}}_i \|^2 ). \label{H_exp}
\end{align}

Define
\begin{equation*}
  \bar{f}_{Ti} = \frac{1}{T} \sum_{t=1}^T f_i ( \mathbf{z}_{it}\tr (\widehat{\bm{\theta}}_i - \bm{\theta}_{i0}) | \mathbf{z}_{it} )
\end{equation*}
and
\begin{equation*}
  \bar{\mathbf{g}}_{Ti} = \bar{f}_{Ti}^{-1} \frac{1}{T} \sum_{t=1}^T f_i ( \mathbf{z}_{it}\tr (\widehat{\bm{\theta}}_i - \bm{\theta}_{i0}) | \mathbf{z}_{it} ) \mathbf{x}_{it}.
\end{equation*}
  Under assumptions~\ref{a:data}-\ref{a:minf} (which imply Lemma~\ref{lem:consistent_est}), for each $i$ these functions converge in probability as $T \rightarrow \infty$ to $\mathbf{g}_i$ and $f_i$ defined respectively in~\eqref{eq:def-gi} and just before.

  Using~\eqref{H_exp} in~\eqref{alpha_expansion}, the definitions just introduced and solving for $\widehat{\alpha}_i^{\ast} - \widehat{\alpha}_i$ we find that for all $i$,
\begin{multline} \label{alpha_i_exp}
  \widehat{\alpha}_i^{\ast} - \widehat{\alpha}_i = -\bar{f}_{Ti}^{-1} \mathbb{H}_{Ti} (\widehat{\bm{\theta}}_i) - \bar{\mathbf{g}}_{Ti}\tr (\widehat{\bm{\beta}}^{\ast} - \widehat{\bm{\beta}}) - \bar{f}_{Ti}^{-1} \mathcal{H}_{Ti}(\widehat{\bm{\theta}}_i^\ast, \widehat{\bm{\theta}}_i) \\
  + \bar{f}_{Ti}^{-1} \mathbb{H}_{Ti} (\widehat{\bm{\theta}}_i^{\ast}) + O_{p^{\ast}}(\| \widehat{\bm{\theta}}^{\ast}_i - \widehat{\bm{\theta}}_i \|^2).
\end{multline}

Similarly write
\begin{equation}
  \mathbb{S}_{nT}^{\ast} (\widehat{\bm{\theta}}^{\ast}) = \mathbb{S}_{nT}^{\ast} (\widehat{\bm{\theta}}) + \left( S^\ast_{nT} (\widehat{\bm{\theta}}^{\ast}) - S^\ast_{nT} (\widehat{\bm{\theta}}) \right) + \mathcal{S}_{nT}^{\ast} (\widehat{\bm{\theta}}^{\ast}, \widehat{\bm{\theta}}) \label{beta_expansion}
\end{equation}
with
\begin{equation*}
  \mathcal{S}_{nT}^{\ast} (\bm{\theta}, \bm{\theta}') := \mathbb{S}_{nT}^{\ast} (\bm{\theta}) - S^\ast_{nT} (\bm{\theta}) - \mathbb{S}_{nT}^{\ast} (\bm{\theta}') + S^\ast_{nT} (\bm{\theta}').
\end{equation*}

Using calculations similar to those for $H_{Ti}(\widehat{\bm{\theta}}_i^{\ast}) - H_{Ti}(\widehat{\bm{\theta}}_i)$, we may rewrite
\begin{multline} \label{S_exp}
  S^\ast_{nT}(\widehat{\bm{\theta}}^{\ast}) - S^\ast_{nT}(\widehat{\bm{\theta}}) = \frac{1}{nT} \sum_{i=1}^n \omega_i \sum_{t=1}^T f_i (\mathbf{z}_{it}\tr (\widehat{\bm{\theta}}_i - \bm{\theta}_{i0}) | \mathbf{z}_{it} ) \mathbf{x}_{it} (\widehat{\alpha}_i^{\ast} - \widehat{\alpha}_i) \\
  + \frac{1}{nT} \sum_{i=1}^n \omega_i \sum_{t=1}^T f_i (\mathbf{z}_{it}\tr (\widehat{\bm{\theta}}_i - \bm{\theta}_{i0}) | \mathbf{z}_{it} ) \mathbf{x}_{it} \mathbf{x}_{it}\tr (\widehat{\bm{\beta}}^{\ast} - \widehat{\bm{\beta}}) + O_{p^{\ast}} \left( \| \widehat{\bm{\beta}}^\ast - \widehat{\bm{\beta}} \|^2 + \sup_i |\widehat{\alpha}_i^\ast - \widehat{\alpha}_i|^2 \right).
\end{multline}

Next let
\begin{align*}
  \widehat{\Gamma}^\ast_{nT} &:= \frac{1}{nT} \sum_{i=1}^n \omega_i \sum_{t=1}^T f_i (\mathbf{z}_{it}\tr (\widehat{\bm{\theta}}_i - \bm{\theta}_{i0}) | \mathbf{z}_{it} ) \mathbf{x}_{it} ( \mathbf{x}_{it} - \bar{\mathbf{g}}_{Ti} )\tr, \\
  \mathbb{H}_{nT}^\ast(\bm{\theta}) &:= \mathbb{S}_{nT}^\ast(\bm{\theta}) - \frac{1}{n} \sum_{i=1}^n \omega_i \bar{\mathbf{g}}_{Ti} \mathbb{H}_{Ti}(\bm{\theta}_i), \\
  H_{nT}^\ast(\bm{\theta}) &:= S_{nT}^\ast(\bm{\theta}) - \frac{1}{n} \sum_{i=1}^n \omega_i \bar{\mathbf{g}}_{Ti} H_{Ti}(\bm{\theta}_i), \\
  \mathcal{H}_{nT}^\ast(\bm{\theta}, \bm{\theta}') &:= \mathbb{H}_{nT}^\ast(\bm{\theta}) - H_{nT}^\ast(\bm{\theta}) - \mathbb{H}_{nT}^\ast(\bm{\theta}') + H_{nT}^\ast(\bm{\theta}').
\end{align*}

Use~\eqref{alpha_i_exp} in~\eqref{S_exp} and then put the resulting expression in~\eqref{beta_expansion}.  Using the above definitions, rearrangement results in
\begin{equation} \label{S_exp_last}
  \widehat{\Gamma}_{nT}^\ast (\widehat{\bm{\beta}}^{\ast} - \widehat{\bm{\beta}}) = -\mathbb{H}_{nT}^\ast(\widehat{\bm{\theta}}) - \mathcal{H}_{nT}^\ast(\widehat{\bm{\theta}}^\ast, \widehat{\bm{\theta}}) + \mathbb{H}_{nT}^\ast(\widehat{\bm{\theta}}^\ast) + O_{p^{\ast}} \left( \| \widehat{\bm{\beta}}^\ast - \widehat{\bm{\beta}} \|^2 + \sup_i |\widehat{\alpha}_i^\ast - \widehat{\alpha}_i|^2 \right).
\end{equation}

Now we determine stochastic orders for all the terms in the above expression.  It can be verified that $\hat{\Gamma}_{nT}^\ast \stackrel{p^\ast}{\rightarrow} \Gamma$, and is nonsingular with probability tending to one.  Lemma 4.6 of \citet{PraestgaardWellner93}, with $a_j = \frac{1}{T} \sum_{t=1}^T (\mathbf{x}_{it} - \bar{\mathbf{g}}_{Ti}) \psi_\tau(\widehat{e}_{it})$ (using Lemma 19.24 of \citet{vanderVaart98} to note that the asymptotic variance of $\sqrt{T} a_j$ is $V$), and $B_j = \omega_i$ implies
\begin{equation} \label{multiplier_bahadur}
  -\sqrt{nT} \mathbb{H}_{nT}^\ast(\widehat{\bm{\theta}}) \dconvs \mathcal{N} \left( \mathbf{0}, V \right).
\end{equation}
Also, Lemma~\ref{lem:subg} above implies that
\begin{equation} \label{qr_soln}
  \mathbb{H}_{nT}^\ast(\widehat{\bm{\theta}}^\ast) = O(1 / T) \; a.s.
\end{equation}
It can also be verified that the calculations in Lemma 7 of \citet{GalvaoGuVolgushev20} using $\widehat{\alpha}_i^\ast, \widehat{\alpha}_i$ in the place of $\widehat{\alpha}_i, \tilde{\alpha}_i$ (noting that here, the equivalent of their $\tilde{\alpha}_i$ is $\widehat{\alpha}_i$) show that
\begin{equation} \label{alpha_order_easy}
  \sup_i | \widehat{\alpha}_i^\ast - \widehat{\alpha}_i | = O_{p^\ast} \left( \|\widehat{\bm{\beta}}^\ast - \widehat{\bm{\beta}}\| + T^{-1} (\log T)^2 \right).
\end{equation}
To discuss the order of $\mathcal{H}_{nT}^\ast(\widehat{\bm{\theta}}^\ast, \widehat{\bm{\theta}})$, define
\begin{equation*}
  \| \mathbb{P}_{Ti} - P_i \|_{\mathcal{G}} = \sup_{g \in \mathcal{G}} \left| \frac{1}{T} \sum_{t=1}^T \left( g(y_{it}, z_{it}) - \E[g(Y_{it}, Z_{it})] \right) \right|,
\end{equation*}
and let $\mathcal{G}_2(\delta)$ be the family of functions defined in Lemma~5 of \citet{GalvaoGuVolgushev20}:
\begin{multline*}
  \mathcal{G}_2(\delta) = \big\{ (y, z) \mapsto a\tr z (I(y \leq z\tr b_1) - I(y \leq z\tr b_2)) I(\|z\| \leq M) : \\
  b_1, b_2 \in \R^{p+1}, \|b_1 - b_2\| \leq \delta, a \in \mathcal{S}^{p+1} \big\}.
\end{multline*}
Let $\omega_i \mathcal{H}_i$ denote the $i$th term in the average making up $\mathcal{H}_{nT}^\ast(\widehat{\bm{\theta}}^\ast, \widehat{\bm{\theta}})$.  Then
\begin{equation*}
  \mathcal{H}_{nT}^\ast(\widehat{\bm{\theta}}^\ast, \widehat{\bm{\theta}}) = O_{p^\ast} \left(  \sup_i |\mathcal{H}_i| \times \frac{1}{n} \sum_{i=1}^n \omega_i \right).
\end{equation*}
We may bound the order of $\mathcal{H}_i$ as follows.  First,
\begin{equation*}
  \sup_i | \mathcal{H}_i | \leq M(1 + f_{\text{max}} / f_{\text{min}}) \sup_i \| \mathbb{P}_{Ti} - P_i \|_{\mathcal{G}_2(\|\widehat{\bm{\beta}}^\ast - \widehat{\bm{\beta}}\| + \sup_i |\widehat{\alpha}_i^\ast - \widehat{\alpha}_i|)}.
\end{equation*}
Lemma~5 of \citet{GalvaoGuVolgushev20} implies that
\begin{multline*}
  \sup_i \| \mathbb{P}_{Ti} - P_i \|_{\mathcal{G}_2(\|\widehat{\bm{\beta}}^\ast - \widehat{\bm{\beta}}\| + \sup_i |\widehat{\alpha}_i^\ast - \widehat{\alpha}_i|)} \\
  = O_{p^\ast} \left( (\|\widehat{\bm{\beta}}^\ast - \widehat{\bm{\beta}}\| + \sup_i |\widehat{\alpha}_i^\ast - \widehat{\alpha}_i|)^{1/2} T^{-1/2} \log T + T^{-1} (\log T)^2 \right).
\end{multline*}
Using~\eqref{alpha_order_easy} we have that
\begin{multline*}
  O_{p^\ast} \left( (\|\widehat{\bm{\beta}}^\ast - \widehat{\bm{\beta}}\| + \sup_i |\widehat{\alpha}_i^\ast - \widehat{\alpha}_i|)^{1/2} T^{-1/2} \log T + T^{-1} (\log T)^2 \right) \\
  = O_{p^\ast} \left( \|\widehat{\bm{\beta}}^\ast - \widehat{\bm{\beta}}\|^{1/2} T^{-1/2} \log T + T^{-1} (\log T)^2 \right).
\end{multline*}
Meanwhile, the average of the $\omega_i$ is bounded in probability by the assumption that $\omega_i$ have a finite second moment, implying that
\begin{equation} \label{hscript_order_easy}
  \mathcal{H}_{nT}^\ast(\widehat{\bm{\theta}}^\ast, \widehat{\bm{\theta}}) = O_{p^\ast} \left( \|\widehat{\bm{\beta}}^\ast - \widehat{\bm{\beta}}\|^{1/2} T^{-1/2} \log T + T^{-1} (\log T)^2 \right).
\end{equation}
Using~\eqref{multiplier_bahadur}--\eqref{hscript_order_easy} in~\eqref{S_exp_last} results in
\begin{equation*}
  \|\widehat{\bm{\beta}}^\ast - \widehat{\bm{\beta}}\| = O_{p^\ast}\left( (nT)^{-1/2} \right) + O_{p^\ast} \left( \|\widehat{\bm{\beta}}^\ast - \widehat{\bm{\beta}}\|^{1/2} T^{-1/2} \log T \right) + O_{p^\ast} \left( T^{-1} (\log T)^2 \right).
\end{equation*}
Then ``fact 1'' from page 198 of \citet{GalvaoGuVolgushev20} implies that
\begin{equation} \label{beta_order}
  \|\widehat{\bm{\beta}}^\ast - \widehat{\bm{\beta}}\| = O_{p^\ast}\left( (nT)^{-1/2} + T^{-1} (\log T)^2 \right).
\end{equation}
Finally, using~\eqref{beta_order} in~\eqref{hscript_order_easy} and~\eqref{qr_soln}--\eqref{hscript_order_easy} in~\eqref{S_exp_last} implies that
\begin{equation*}
  \widehat{\bm{\beta}}^\ast - \widehat{\bm{\beta}} = - (\widehat{\Gamma}_{nT}^\ast)^{-1} \mathbb{H}_{nT}^\ast(\widehat{\bm{\theta}}) + O_{p^\ast} \left( ((nT)^{-1/4} + T^{-1/2} \log T) T^{-1/2} \log T \right),
\end{equation*}
so the rate restriction on $n, T$ and~\eqref{multiplier_bahadur} imply the result.
\end{proof}

\begin{proof}[Proof of Theorem~\ref{thm:consistent}]
The proof follows directly from Lemma \ref{lem:Anormal} above and Theorem 3 in \citet{GalvaoGuVolgushev20}.
\end{proof}

\begin{proof}[Proof of Theorem~\ref{thm:consistent_var}]
  Recall the definition of $\widehat{\Sigma}^\ast$ in~\eqref{eq:boot_var}.  We show the existence of a $q := 2 + \epsilon$ moment (where $\epsilon > 0$) of $\sqrt{nT}\|\widehat{\bm{\beta}}^\ast - \widehat{\bm{\beta}}\|$ uniformly in $n, T$, since that implies uniform square integrability of the sequence of $\widehat{\Sigma}^\ast$, and combined with the weak convergence shown in Lemma~\ref{lem:Anormal}, implies the result \citep[Corollary 7, p.277]{ChowTeicher97}.

  Follow the main steps in the expansions at the beginning of proof of Lemma~\ref{lem:Anormal} using the notation defined there, but record the remainder terms in detail.  Briefly, expression~\eqref{H_exp} may be rewritten
  \begin{align*}
    H_{Ti} (\widehat{\bm{\theta}}_i^{\ast}) - H_{Ti} (\widehat{\bm{\theta}}_i) &= \frac{1}{T} \sum_{t=1}^T f_i ( \mathbf{z}_{it}\tr (\widehat{\bm{\theta}}_i - \bm{\theta}_{i0}) | \mathbf{z}_{it} ) (\widehat{\alpha}_i^{\ast} - \widehat{\alpha}_{i}) \notag \\
    {} &\phantom{=} \quad + \frac{1}{T} \sum_{t=1}^T f_i ( \mathbf{z}_{it}\tr (\widehat{\bm{\theta}}_i - \bm{\theta}_{i0}) | \mathbf{z}_{it} ) \mathbf{x}_{it}\tr (\widehat{\bm{\beta}}^{\ast} - \widehat{\bm{\beta}}) + R_{1Ti}^\ast,
  \end{align*}
  where for each $i$,
  \begin{equation*}
    R_{1Ti}^\ast = \frac{1}{T} \sum_{t=1}^T \left( F_i(\mathbf{z}_{it}\tr (\widehat{\bm{\theta}}_i^\ast - \bm{\theta}_{i0}) | \mathbf{z}_{it}) - F_i(\mathbf{z}_{it}\tr (\widehat{\bm{\theta}}_i - \bm{\theta}_{i0}) | \mathbf{z}_{it}) - f_i(\mathbf{z}_{it}\tr (\widehat{\bm{\theta}}_i - \bm{\theta}_{i0}) | \mathbf{z}_{it}) \mathbf{z}_{it}\tr (\widehat{\bm{\theta}}_i^\ast - \widehat{\bm{\theta}}_i) \right).
  \end{equation*}
  Similarly, \eqref{S_exp} can be rewritten
  \begin{multline*}
    S^\ast_{nT}(\widehat{\bm{\theta}}^{\ast}) - S^\ast_{nT}(\widehat{\bm{\theta}}) = \frac{1}{nT} \sum_{i=1}^n \omega_i \sum_{t=1}^T f_i (\mathbf{z}_{it}\tr (\widehat{\bm{\theta}}_i - \bm{\theta}_{i0}) | \mathbf{z}_{it} ) \mathbf{x}_{it} (\widehat{\alpha}_i^{\ast} - \widehat{\alpha}_i) \\
    + \frac{1}{nT} \sum_{i=1}^n \omega_i \sum_{t=1}^T f_i (\mathbf{z}_{it}\tr (\widehat{\bm{\theta}}_i - \bm{\theta}_{i0}) | \mathbf{z}_{it} ) \mathbf{x}_{it} \mathbf{x}_{it}\tr (\widehat{\bm{\beta}}^{\ast} - \widehat{\bm{\beta}}) + R_{2nT}^\ast,
  \end{multline*}
  with
  \begin{multline*}
    R_{2nT}^\ast = \frac{1}{nT} \sum_{i=1}^n \omega_i \sum_{t=1}^T \mathbf{x}_{it} \Big( F_i(\mathbf{z}_{it}\tr (\widehat{\bm{\theta}}_i^\ast - \bm{\theta}_{i0}) | \mathbf{z}_{it}) - F_i(\mathbf{z}_{it}\tr (\widehat{\bm{\theta}}_i - \bm{\theta}_{i0}) | \mathbf{z}_{it}) \\
    - f_i(\mathbf{z}_{it}\tr (\widehat{\bm{\theta}}_i - \bm{\theta}_{i0}) | \mathbf{z}_{it}) \mathbf{z}_{it}\tr (\widehat{\bm{\theta}}_i^\ast - \widehat{\bm{\theta}}_i) \Big).
  \end{multline*}
  Finally, let
  \begin{align*}
    R_{nT}^\ast &= R_{2nT}^\ast - \frac{1}{n} \sum_{i=1}^n \omega_i \bar{\mathbf{g}}_{Ti} R_{1Ti}^\ast \\
    {} &= \frac{1}{nT} \sum_{i=1}^n \omega_i \sum_{t=1}^T (\mathbf{x}_{it} - \bar{\mathbf{g}}_{Ti}) \Big( F_i(\mathbf{z}_{it}\tr (\widehat{\bm{\theta}}_i^\ast - \bm{\theta}_{i0}) | \mathbf{z}_{it}) - F_i(\mathbf{z}_{it}\tr (\widehat{\bm{\theta}}_i - \bm{\theta}_{i0}) | \mathbf{z}_{it}) \\
    {} &\phantom{=} \qquad - f_i(\mathbf{z}_{it}\tr (\widehat{\bm{\theta}}_i - \bm{\theta}_{i0}) | \mathbf{z}_{it}) \mathbf{z}_{it}\tr (\widehat{\bm{\theta}}_i^\ast - \widehat{\bm{\theta}}_i) \Big).
  \end{align*}

  We may put these expressions into equation~\eqref{S_exp_last} from the proof of Theorem~\ref{lem:Anormal} to find that
  \begin{multline} \label{last_with_remainder}
    \E^\ast \left[ (nT)^{q/2} \left\| (\widehat{\bm{\beta}}^{\ast} - \widehat{\bm{\beta}}) \right\|^q \right] \\
    = \E^\ast \left[ (nT)^{q/2} \left\| (\widehat{\Gamma}_{nT}^\ast)^{-1} \left( -\mathbb{H}_{nT}^\ast(\widehat{\bm{\theta}}) - \mathcal{H}_{nT}^\ast(\widehat{\bm{\theta}}^\ast, \widehat{\bm{\theta}}) + \mathbb{H}_{nT}^\ast(\widehat{\bm{\theta}}^\ast) + R_{nT}^\ast \right) \right\|^q \right].
  \end{multline}

  We will show that the right-hand side of~\eqref{S_exp_last} is bounded in probability by considering individual terms in the sum.  For any $\bm{\theta}$,
  \begin{align*}
    \left\| \mathbb{H}_{nT}^\ast(\bm{\theta}) \right\| &= \left\| \frac{1}{nT} \sum_{i=1}^n \omega_i \sum_{t=1}^T (\mathbf{x}_{it} - \bar{\mathbf{g}}_{Ti}) \psi_\tau \left( e_{it} - \mathbf{z}_{it}\tr (\bm{\theta}_i - \bm{\theta}_{i0}) \right) \right\| \\
    {} &\leq \left\| \frac{1}{nT} \sum_{i=1}^n \omega_i \sum_{t=1}^T (\mathbf{x}_{it} - \bar{\mathbf{g}}_{Ti}) \right\|.
  \end{align*}
  Similarly,
  \begin{equation*}
    \left\| H_{nT}^\ast(\bm{\theta}) \right\| \leq \left\| \frac{1}{nT} \sum_{i=1}^n \omega_i \sum_{t=1}^T (\mathbf{x}_{it} - \bar{\mathbf{g}}_{Ti}) \right\|.
  \end{equation*}
  Note that $\mathcal{H}_{nT}^\ast$ is a sum of several such terms.  Lastly, note that for each $i$,
  \begin{equation*}
    F_i(\mathbf{z}_{it}\tr (\widehat{\bm{\theta}}_i^\ast - \bm{\theta}_{i0}) | \mathbf{z}_{it}) - F_i(\mathbf{z}_{it}\tr (\widehat{\bm{\theta}}_i - \bm{\theta}_{i0}) | \mathbf{z}_{it}) = f_i(\mathbf{z}_{it}\tr (\bar{\bm{\theta}}_i - \bm{\theta}_{i0}) | \mathbf{z}_{it}) \mathbf{z}_{it}\tr (\widehat{\bm{\theta}}_i^\ast - \widehat{\bm{\theta}}_i),
  \end{equation*}
  where $\bar{\bm{\theta}}_i$ is between $\widehat{\bm{\theta}}_i^\ast$ and $\widehat{\bm{\theta}}_i$.  Under assumption~\ref{a:boundedZ}, there are some $\bar{A}, \bar{B}$ such that  $|\alpha_i| \leq \bar{A}$ for all $i$ and $\|\bm{\beta}\| \leq \bar{B}$.  Therefore
  \begin{align*}
    \| R_{nT}^\ast \| &= \left\| \frac{1}{nT} \sum_{i=1}^n \omega_i \sum_{t=1}^T (\mathbf{x}_{it} - \bar{\mathbf{g}}_{Ti}) \left( f_i(\mathbf{z}_{it}\tr (\bar{\bm{\theta}}_i - \bm{\theta}_{i0}) | \mathbf{z}_{it}) - f_i(\mathbf{z}_{it}\tr (\widehat{\bm{\theta}}_i - \bm{\theta}_{i0}) | \mathbf{z}_{it}) \right) \mathbf{z}_{it}\tr (\widehat{\bm{\theta}}_i^\ast - \widehat{\bm{\theta}}_i) \right\| \\
    {} &\leq 2f_{\text{max}}(M+1) (\bar{A} + \bar{B}) \left\| \frac{1}{nT} \sum_{i=1}^n \omega_i \sum_{t=1}^T (\mathbf{x}_{it} - \bar{\mathbf{g}}_{Ti}) \right\|.
  \end{align*}

  Therefore the $c_r$ inequality and the above bounds imply that the right-hand side of~\eqref{last_with_remainder} is finite if
  \begin{equation} \label{qmoment_rhs_bound}
    \E^\ast \left[ \left\| \frac{1}{\sqrt{nT}} \sum_{i=1}^n \omega_i \sum_{t=1}^T (\mathbf{x}_{it} - \bar{\mathbf{g}}_{Ti}) \right\|^q \right] < \infty.
  \end{equation}

  To simplify notation, let $\mu_p := \E^\ast[\omega_i^p]$.  These moments exist for $0 < p < q$ by Assumption~\ref{a:moments}.  Using Rosenthal's inequality \citep{Hansen15} for the first inequality, there exists $C_q < \infty$ such that
  \begin{align*}
    \E^\ast \Bigg[ \bigg\| \frac{1}{\sqrt{nT}} \sum_{i=1}^n &\omega_i \sum_{t=1}^T (\mathbf{x}_{it} - \bar{\mathbf{g}}_{Ti}) \Bigg\|^q \Bigg] \\
    {} &\leq \frac{C_q}{(nT)^{q/2}} \left( \left( \sum_{i=1}^n \E^\ast \left[ \left\| \omega_i \sum_{t=1}^T (\mathbf{x}_{it} - \bar{\mathbf{g}}_{Ti}) \right\|^2 \right] \right)^{q/2} + \sum_{i=1}^n \E^\ast \left[ \left\| \omega_i \sum_{t=1}^T (\mathbf{x}_{it} - \bar{\mathbf{g}}_{Ti}) \right\|^q \right] \right) \\
    {} &= \frac{C_q}{(nT)^{q/2}} \left( \left( \sum_{i=1}^n \mu_2 \left\| \sum_{t=1}^T (\mathbf{x}_{it} - \bar{\mathbf{g}}_{Ti}) \right\|^2 \right)^{q/2} + \sum_{i=1}^n \mu_q \left\| \sum_{t=1}^T (\mathbf{x}_{it} - \bar{\mathbf{g}}_{Ti}) \right\|^q \right).
  \end{align*}
  This last line is the sum of two terms.  Under Assumption~\ref{a:moments} and the definition of $\bar{\mathbf{g}}_{Ti}$, which ensures that it is at most $f_{\text{max}} / f_{\text{min}}$ the scale of $\frac{1}{T} \sum_t \mathbf{x}_{it}$,
  \begin{equation*}
    \E \left[ \left( \frac{1}{nT} \sum_{i=1}^n \left\| \sum_{t=1}^T (\mathbf{x}_{it} - \bar{\mathbf{g}}_{Ti}) \right\|^2 \right)^{q/2} \right] < \infty,
  \end{equation*}
  so this term converges in probability to a finite limit.  For the second term, note that
  \begin{equation*}
    \frac{1}{(nT)^{q/2}} \sum_{i=1}^n \left\| \sum_{t=1}^T (\mathbf{x}_{it} - \bar{\mathbf{g}}_{Ti}) \right\|^q \leq \frac{1}{n} \sum_{i=1}^n \left\| \frac{1}{\sqrt{T}} \sum_{t=1}^T (\mathbf{x}_{it} - \bar{\mathbf{g}}_{Ti}) \right\|^q,
  \end{equation*}
  which also has bounded expectation, and thus converges in probability to a finite limit.  Therefore these two terms are $q$-integrable uniformly in $n, T$, implying that~\eqref{qmoment_rhs_bound} is finite and the left-hand side of~\eqref{last_with_remainder} is finite as well.  The convergence of $\widehat{\Gamma}_{nT}^\ast$ in probability, conditional on the observations, to the nonsingular $\Gamma$ implies the result.
  \end{proof}

\section{List of Countries}\label{appB}

\begin{footnotesize}
\begin{longtable} {ccccc}
\caption{List of OECD countries}
\label{tblistOECD}\\
 \hline
Australia	&	Denmark	&	Japan	&	New Zealand	&	Switzerland	\\
Austria	&	Finland	&	South Korea 	&	Norway	&	Turkey	\\
Belgium	&	France	&	Luxembourg 	&	Portugal	&	United Kingdom	\\
Canada	&	Ireland	&	Mexico	&	Spain	&	 United States 	\\
Chile	&	Italy	&	The Netherlands 	&	Sweden	&		\\
\hline
\end{longtable}
\end{footnotesize}

\begin{footnotesize}
\begin{longtable} {ccccc}
\caption{List of non-OECD countries}
\label{tblistNonOECD}\\
 \hline
Argentina	&	Cyprus	&	Hong Kong of China 	&	Pakistan	&	South Africa	\\
Bahamas	&	Dominican Republic	&	Israel	&	Panama	&	Sri Lanka	\\
Belarus	&	Ecuador	&	Jamaica	&	Paraguay	&	Thailand	\\
Brazil 	&	El Salvador	&	Malaysia	&	Peru	&	Uruguay	\\
China	&	Gabon	&	Nepal	&	Philippines	&		\\
Colombia	&	Guatemala 	&	Nicaragua	&	Senegal  	&		\\
Costa Rica	&	Honduras	&	Nigeria	&	Singapore	&		\\
 \hline
\end{longtable}
\end{footnotesize}

\newpage

\bibliographystyle{econometrica}
\bibliography{pqrboot}

\end{document}